\documentclass[10pt]{article}
\usepackage[left=0.8in,top=0.8in,right=0.8in,bottom=0.8in]{geometry}

\usepackage{graphicx}
\usepackage{amsmath}
\usepackage{amssymb}
\usepackage{amsthm}
\usepackage{latexsym}
\usepackage{bm}
\usepackage{array,supertabular}
\usepackage{color}
\usepackage{framed}
\usepackage{setspace}
\usepackage{fancyhdr}
\usepackage{stmaryrd}
\usepackage{mathrsfs}
\usepackage{mdframed}
\usepackage{url}
\usepackage{multirow}
\usepackage{algorithm}
\usepackage{algpseudocode}
\usepackage{pifont}
\usepackage{enumerate}
\newcolumntype{P}[1]{>{\centering\arraybackslash}p{#1}}

\SetSymbolFont{stmry}{bold}{U}{stmry}{m}{n}

\newtheorem{remark}{Remark}

\numberwithin{equation}{section}

\begin{document}
\title{Modeling fibrous tissue in vascular fluid-structure interaction: a morphology-based pipeline and biomechanical significance}
\author{Yujie Sun$^{\textup{ a}}$, Jiayi Huang$^{\textup{ a}}$, Qingshuang Lu$^{\textup{ a}}$, Xinhai Yue$^{\textup{ a}}$, Xuanming Huang$^{\textup{ a}}$, \\
Wei He$^{\textup{ b}}$, Yun Shi$^{\textup{ b}}$, Ju Liu$^{\textup{ a,*}}$ \\
$^a$ \textit{\small Department of Mechanics and Aerospace Engineering,}\\
\textit{\small Southern University of Science and Technology,}\\
\textit{\small 1088 Xueyuan Avenue, Shenzhen, Guangdong 518055, China}\\
$^b$ \textit{\small Institute of Vascular Surgery, Department of Vascular Surgery, Zhongshan Hospital, Fudan University,}\\
\textit{\small 180 Fenglin Road, Shanghai 200032, China}\\
$^{*}$ \small \textit{E-mail address:} liuj36@sustech.edu.cn
}
\date{}
\maketitle

\section*{Abstract}
Modeling fibrous tissue for vascular fluid-structure interaction analysis poses significant challenges due to the lack of effective tools for preparing simulation data from medical images. This limitation hinders the physiologically realistic modeling of vasculature and its use in clinical settings. Leveraging an established lumen modeling strategy, we propose a comprehensive pipeline for generating thick-walled artery models. A specialized mesh generation procedure is developed to ensure mesh continuity across the lumen and wall interface. Exploiting the centerline information, a series of procedures are introduced for generating local basis vectors within the arterial wall. The procedures are tailored to handle thick-walled tissues where basis vectors may exhibit transmural variations. Additionally, we propose methods for accurately identifying the centerline in multi-branched vessels and bifurcating regions. These modeling approaches are algorithmically implementable, rendering them readily integrable into mainstream cardiovascular modeling software. The developed fiber generation method is evaluated against the strategy using linear elastic analysis, demonstrating that the proposed approach yields satisfactory fiber definitions in the considered benchmark. Finally, we examine the impact of anisotropic arterial wall models on the vascular fluid-structure interaction analysis through numerical examples, employing the neo-Hookean model for comparative purposes. The first case involves an idealized curved geometry, while the second studies an image-based abdominal aorta model. Our numerical results reveal that the deformation and stress distribution are critically related to the constitutive model of the wall, whereas hemodynamic factors are less sensitive to the wall model. This work paves the way for more accurate image-based vascular modeling and enhances the prediction of arterial behavior under physiologically realistic conditions.

\vspace{5mm}

\noindent \textbf{Keywords:} Fluid-structure interaction, Image-based modeling, Patient-specific simulation, Anisotropic material model,  Vascular biomechanics

\section{Introduction}
Over the years, image-based computational fluid dynamics (CFD) and fluid-structure interaction (FSI) gradually get matured and offer a non-invasive approach for quantifying hemodynamic factors, including pressure, wall shear stress (WSS), fractional flow reserve, etc \cite{Hirschhorn2020,Taylor2023,Updegrove2017}. These factors are crucial for understanding disease progression, making surgical plans, and designing medical devices \cite{Taylor2013}. As a more physiologically realistic model for vasculature, FSI has been demonstrated to be superior in assessing WSS and related biomarkers \cite{Liu2020,Lopes2019,Reymond2013}, characterizing pulse wave propagation \cite{Lan2023,Mobadersany2023}, and most importantly, analyzing stresses of the tissue wall \cite{Bazilevs2010}. The latter plays a critical role in a variety of clinical scenarios, such as risk assessment of aneurysms \cite{Sherifova2019}, evaluation of the impact of balloon angioplasty on vascular tissue \cite{Bukac2019}, etc.
 
Despite the encouraging progress in image-based modeling, we witness a dichotomy between the fields of vascular FSI and vascular constitutive modeling. Most existing vascular FSI studies neglect physiological details that are discovered and quantified in biomechanical modeling. For example, \textit{isotropic} hyperelastic models are still widely used to characterize the mechanical behavior of arterial walls in vascular FSI investigations \cite{Baeumler2020,Bruneau2023,Bukac2019,Liu2020,Lopes2019,Moerman2022,Souche2022,Suito2014}. In the meantime, the detailed constitutive models that incorporate biomechanics and mechanobiology information have been actively developed \cite{Humphrey2021,Sherifova2019,Pukaluk2024}. Some recent FSI studies, albeit based on idealized geometries, have revealed the significance of anisotropy, especially in transmural stress distribution \cite{Balzani2016,Balzani2023,Tricerri2015}. We view the one major hurdle in the field is the lack of robust tools in characterizing the tissue wall based on physiological facts and image data. This thus becomes the primary research thrust of our work.

\subsection{Physiological background and constitutive modeling}
In histology, like most biological tissues, arteries are composed of extracellular matrix (e.g., elastin and collagen), cells (e.g., endothelial and smooth muscle cells), and water. Specifically, the collagen fibers and smooth muscles provide significant stiffness when being stretched and are embedded in groundmatrix which is isotropic and much softer. This composition endows the tissue with directional and nonlinear attributes. Due to the abundant content of water, most tissues behave in a nearly incompressible manner. The arterial wall consists of three layers, characterized by distinct thickness, components, and mechanical properties \cite{Gasser2006,Holzapfel2005}. For example, the collagen fibers in the media form two helically oriented groups, while the orientations in the other layers are more dispersed.

Informed from physiological knowledge and biomechanical measurements, constitutive models for the artery have been evolved to accurately capture physiological details \cite{Dalbosco2024}. Exponential function was introduced to characterize the strain hardening due to collagen fibers \cite{Chagnon2015}. The anisotropy is typically modeled through invariants based on structural tensors \cite{Holzapfel2000a}. The composite nature of the artery is taken into account by distinguishing the non-collagenous groundmatrix and the collagen fibers, which are described by isotropic and anisotropic hyperelastic models, respectively. One representative work was proposed by Holzapfel, Gasser, and Ogden  \cite{Holzapfel2000}, which invokes three parameters for a mechanically homogeneous layer. Through adopting different material parameters, that model can be readily converted to characterize the multi-layer structure of arterial walls. A significant improvement was made by taking the fiber dispersion into account, and it is currently known as the GOH model \cite{Gasser2006}. Further improvement involves the exclusion of fiber contribution under compression \cite{Holzapfel2015} and differentiating the dispersion level in the in-plane and out-of-plane directions \cite{Holzapfel2015a}. We also mention that the well-possedness of a variety of anisotropic arterial tissue models has been confirmed \cite{Balzani2006}, paving the way towards reliable analysis.

\subsection{Image-based modeling}
A mature pipeline exists for preparing the necessary image and physiological data for hemodynamic analysis, with the procedures integrated into finite element analysis software \cite{vmtk-website,Arthurs2021,Updegrove2017}. An underlying concept is to construct the vessel wall as two-dimensional (2D) cross-sectional profiles swept along vessel pathlines. Each cross-section profile is generated by 2D image segmentation and is geometrically represented as closed curves. Through a lofting procedure along the pathline, the curves are joined as a NURBS surface that encloses the lumen of interest. Boolean and smooth operations are necessary to join different luminal branches together and remove non-physiological artifacts from the joined surfaces \cite{opencascade-reference,Updegrove2016}. The eventual geometric model constitutes a boundary representation (B-Rep) of the segmented region, which can be discretized by a mesh generator for simulation purposes. The above-described procedure is typical and relatively mature for subject-specific vascular simulation, despite ongoing progress towards three-dimensional (3D) segmentation, which may eventually bypass the above procedure entirely \cite{Antiga2008,Updegrove2017}.

There are a few critical issues that limit the scope of applicability of the aforementioned modeling strategy. Although the geometric model created is sometimes termed as ``solid model", the word ``solid" is a CAD terminology and has nothing to do with the solid tissue. In fact, the most widely adopted imaging modalities, such as computed tomography angiography and magnetic resonance angiography, cannot identify the vascular and surrounding tissues due to their lack of tissue contrast. The consequent modeled region is the lumen where blood flows. Although people leveraged membrane \cite{Figueroa2006} or shell theories \cite{Nama2020} to enable FSI simulations without invoking a 3D arterial tissue model, we need to be aware that the vessel wall thickness can be about 10\% of the lumen diameter \cite{Caro2012,Humphrey2002}, rendering the thin-wall assumption questionable. Therefore, generating a vessel wall model is compulsory for vascular FSI. One approach for wall modeling relies on the boundary layer (BL) mesh generation algorithm. By switching the layer growth direction to extrude outward from the vessel wall, the generation of volumetric wall meshes becomes feasible and maintains mesh continuity across the fluid-solid interface \cite{Marchandise2013,Raut2015,Wu2022}. Additionally, by setting the extrusion thickness as a fraction of the local vessel radius, variable wall thickness can be achieved \cite{Marchandise2013}. This method works well for geometrically simple vascular structures, such as those with minimal bifurcations and relatively uniform luminal diameters. However, for arterial trees with complex geometric shapes, self-intersection during the extrusion process becomes unavoidable. Advanced BL mesh generation algorithms (such as smoothing, contraction, trimming, etc.) may automatically address the self-intersection issues. When using the algorithms to generate the arterial wall geometry, they may result in physiologically unrealistic geometries of the vessel wall.

Also, the anisotropic constitutive model requires the definition of the fiber orientation at points in the tissue. This is by no means a trivial task for realistic arterial models. Several strategies have been proposed in the literature, each with pros and cons. The first is morphology-based and generates local basis vectors pointed in the circumferential, radial, and axial directions first. The centerline \cite{Marchandise2013,Roy2014} or the tangential planes \cite{Kiousis2009} from the geometric model are utilized for the generation of local basis vectors. With the knowledge of the angle between the fiber direction and the local basis vectors, the fiber orientations are then explicitly defined. This approach has been applied to investigating carotid bifurcation \cite{Kiousis2009} and abdominal aortic aneurysm \cite{Roy2014}. Yet, the considered geometry was relatively simple, and the algorithm needs to be refined to handle multi-branched vessels. In \cite{Alastrue2010}, a static linear elastic analysis was performed to extract the principal stress directions, which are assumed to align with the local radial, circumferential, and axial directions. An even simpler method is based on the Laplace equation with prescribed boundary conditions \cite{Misiulis2019}. The local basis vectors for the mean fiber directions are then derived from the gradient of the resulting scalar field. The above approaches rely on the solution of differential equations, which necessitates a proper boundary condition setting.

The second approach assumes that the fibers can realign and reorient in response to external mechanical stimuli with the goal of enhancing theirs load-bearing capacity. In \cite{Hariton2007}, a biomechanically-driven fiber remodeling method was proposed, in which fiber directions are iteratively updated according to the external load applied. In specific, the fibers are repositioned within a plane spanned by the two maximum principal stress directions, with the angle depending on the ratio of the magnitudes of the two principal stresses. This approach has been applied in a study of carotid bifurcation \cite{Hariton2007b}. In a different approach, a theory was developed to investigate the remodeling of continuously distributed collagen fibers in soft connective tissues, relating the fiber orientation and the mean fiber stretch to the macroscopic stresses within the tissue \cite{Driessen2003}.

\subsection{Contribution and organization of this article}
This contribution of this work involves the following aspects. First, we develop an image-based modeling and mesh generation pipeline to generate high-quality subject-specific meshes suitable for FSI analysis. Instead of manipulating the wall surface mesh, we exploit CAD and polygon mesh processing tools \cite{opencascade-reference,Botsch2010} to deliver a flexible and robust way of model editing. With both the interior and exterior surface representation obtained, we may determine the B-Reps of the lumen as well as the vascular tissue by closing the planar faces at the inlet and outlets, leading to a discrete mesh suitable for FSI analysis \cite{Si2015,Geuzaine2009}. This pipeline is detailed in Section \ref{sec:imaged-based_modeling_mesh_generation}.

Second, we propose a morphology-based approach to define the fiber architecture within the vascular wall in Section \ref{sec:Local_basis}. Our approach relies on the centerline and is refined for complex arterial trees. Compared to the method based on solving partial differential equations, our approach yields a higher quality result near a bifurcation region, as demonstrated in Section \ref{sec:Evaluation_local_basis}. Based on the local basis vectors, the anisotropic constitutive model for arteries can be implemented for subject-specific geometries, which is discussed in Section \ref{sec:constitutive-model}. At the end of Section \ref{sec:methodology}, we present the procedures for generating the prestress within the tissue, which serves as a necessary initialization step for vascular FSI simulations. 

Third, we apply the fiber-reinforced anisotropic constitutive model to investigate the vascular FSI problem. In Section \ref{sec:Curved_tube_FSI}, we consider a simulation of an idealized curved geometry designed in \cite{Balzani2016,Balzani2023} and analyze the impact of the fiber angle and dispersion on the FSI results. After that, we perform an FSI analysis of a patient-specific abdominal aortic model in Section \ref{sec:AA model FSI}, validating the model by comparing with in vivo data and corroborating the observations from the prior idealized geometry. Conclusive remarks are made in Section \ref{sec:conclusion}.

\section{Methodology}
\label{sec:methodology}
\subsection{Image-based modeling pipeline for vascular lumen and tissue}
\label{sec:imaged-based_modeling_mesh_generation}

\subsubsection{Lumen modeling}
\label{sec:lumen-modeling}
In the following, we provide a brief overview of the vascular lumen generation procedures \cite{Arthurs2021,Updegrove2017} that are closely related to the tissue modeling.

\paragraph{Acquisition, preprocessing, and visualization of medical ismages}
After acquiring medical images, they undergo a preprocessing procedure, involving essential steps such as noise reduction, image smoothing, histogram equalization, etc. The objective is to improve the accuracy of subsequent analysis by minimizing artifacts and enhancing contrast. After that, 2D slices from the images are stacked to form a volumetric representation, and volume rendering techniques are employed to visualize anatomical structures in three dimensions. %This procedure is instrumental in the subsequent vascular path planning.

\begin{figure}
\begin{center}
\includegraphics[angle=0, trim=0 0 20 0, clip=true, scale = 0.4]{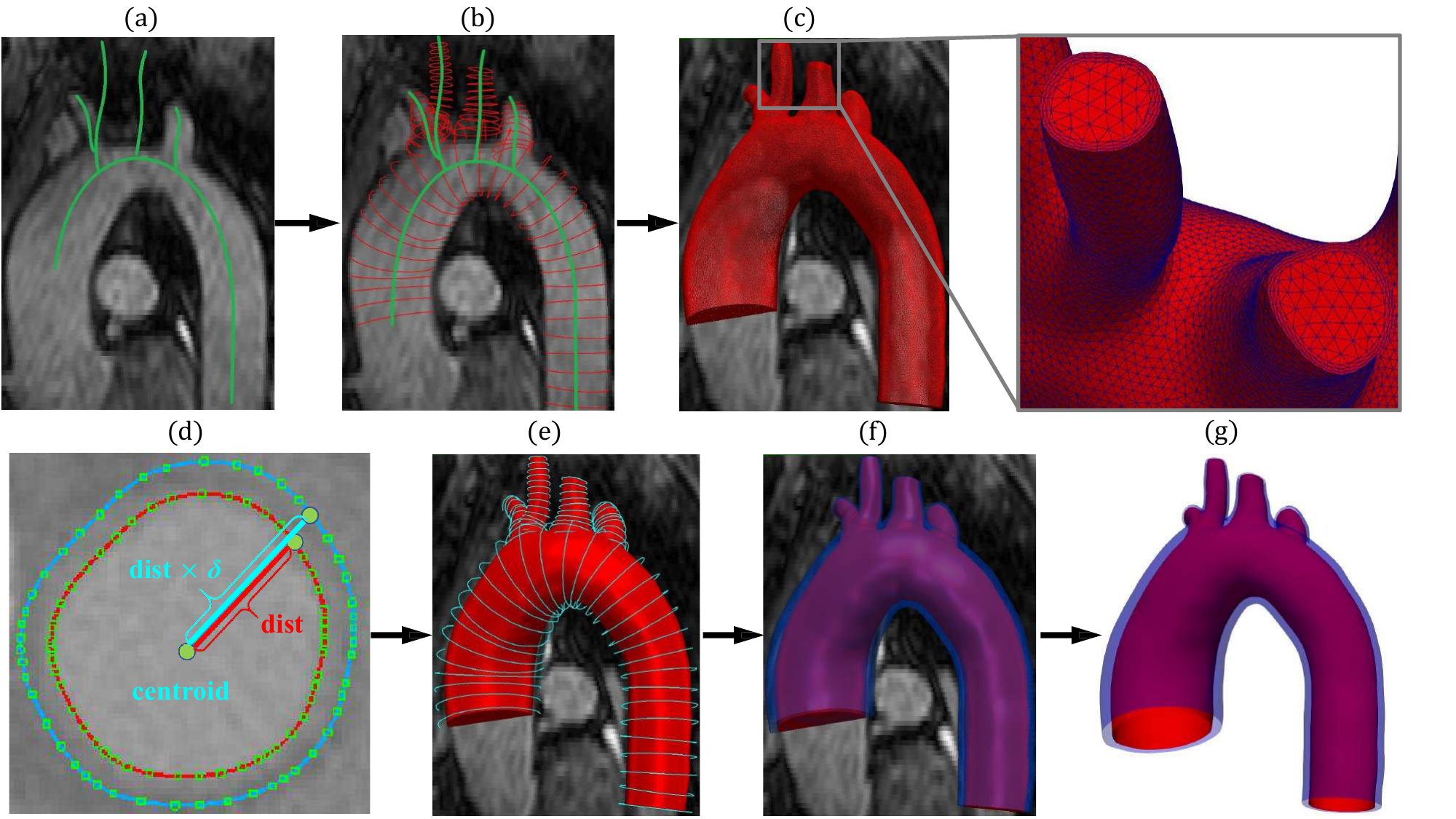}
\end{center}
\caption{(a) Creating pathlines along the vessels of interest; (b) creating 2D segmentations along each pathline; (c) constructing a geometric model and BL mesh; (d) generating the contour points for the exterior surface based on the centroid and the thickness-to-radius ratio $\delta$; (e) creating the closed B-spline curves along each pathline for the tissue exterior surface; (f) constructing the exterior wall surface through lofting; (g) generating the geometric model of the vascular tissue by combining the exterior wall surface with the lumen wall surface.}
\label{fig:mesh-generation-pipeline1}
\end{figure}

\paragraph{2D segmentation along the pathline}
Segmentation divides medical images into sets of pixels to locate areas and boundaries of interest. In this study, 2D image segmentation is employed to extract the geometric shapes of blood vessels, providing essential data for subsequent vascular modeling. We identify points close to the center of the vascular lumen to outline the trajectory of the vessels, and those points are termed as \textit{path points}. Pathlines can then be generated by interpolating the path points, and the planes perpendicular to the pathlines accurately delineate areas apt for segmentation (Figure \ref{fig:mesh-generation-pipeline1} (a)). On each plane, a contour can be generated with extraction methods including level-set \cite{Chan2001,Vese2002} or threshold techniques \cite{Kang2009,Senthilkumaran2016} (Figure \ref{fig:mesh-generation-pipeline1} (b)). %It is worth noting that if the path points are not chosen sufficiently close to the lumen center, the interpolated vascular pathlines may not accurately represent the true curvature of the vessels, causing it difficult to accurately extract image pixel intensities in the 2D profiles of vessels perpendicular to the path.

\paragraph{Lumen surface generation}
Based on the contour generated on each 2D plane, the geometric description of the luminal surface can be obtained through the following procedures. 
\begin{enumerate}
	\item The segmented contour on each plane is represented by a series of contiguous line segments. This contour can be refined into a smooth, closed B-spline curve interpolated using a series of \textit{contour points}, which are sampled from the original contour. Degree elevation and knot insertion are invoked to ensure compatibility of the B-spline curves of different contours.
	\item Lofting along the pathline of the vessel generates a continuous and smooth B-spline surface which seamlessly connects the compatible B-spline curves of each branch \cite{opencascade-reference}. This process generates a tubular geometry that describes the 3D surface of the lumen.
	\item Convert the obtained luminal surface into a discrete mesh. The purpose is to facilitate the later centerline extraction and local smoothing. The conversion can be achieved by the advancing front or Delaunay triangulation techniques \cite{Marchandise2013,Si2015}.
	\item Boolean operations are performed to combine multiple surfaces, forming a complete geometric model of the multi-branched vascular system \cite{opencascade-reference,Updegrove2016}.
	\item It is necessary to employ smoothing techniques to eliminate non-physiological features that arise near the bifurcation location, particularly when there are sharp variations in the vessel radii of different branches. The commonly used smoothing methods include the Laplacian or Taubin methods \cite{Botsch2010}.
\end{enumerate}
The above procedure allows one to obtain a physiologically realistic discrete surface of the vascular lumen, and it has been integrated into open-source vascular modeling software \cite{Arthurs2021,Updegrove2017}. Despite that, the modeling of arterial walls is absent from those software. In the following, we propose a suite of techniques for the modeling of vascular tissue, built upon the lumen surface construction techniques.

\subsubsection{Tissue modeling and mesh generation}
\label{sec:tissue-modeling-meshing}
Based on whether the wall can be identified, the existing imaging modality can be categorized into two groups. The first group involves the computed tomography angiography and magnetic resonance angiography, both of which are widely used nowadays in pathology evaluations. A major limit is their lack of tissue contrast, making it difficult to identify the vascular and surrounding organs. The second group encompasses intravascular ultrasound, optical coherence tomography \cite{Huang2023}, and black-blood magnetic resonance imaging (MRI) \cite{Antiga2008a}, etc. Although they may provide detailed information about the vessel wall and allow direct segmentation, there are critical shortcomings that limit their availability, such as a restricted field of imaging for large vessels, susceptibility to motion artifacts, etc. We assume that the angiographic data does not contain vessel wall information, and the proposed procedures enable the preparation of a vascular tissue model based on most existing imaging modalities.

\paragraph{Tissue surface creation}
We propose the following procedures for constructing the surface representation of the vascular wall. On each plane for 2D segmentation, the arithmetic mean of all contour points defines the \textit{centroid} of the lumen at this cross-section. We then generate a set of new contour points for the exterior wall surface by demanding them to be collinear with their corresponding luminal contour points and the centroid. The distance from the new contour points to the centroid is scaled according to the thickness-to-radius ratio $\delta$. The value of $\delta$ can be chosen according to physiological facts \cite{Caro2012,Humphrey2002}. Alternatively, its value can also be adjusted on each plane to reflect local pathological characteristics. By interpolating the newly generated contour points, we obtain the closed B-spline curve delineating the exterior wall surface (Figure \ref{fig:mesh-generation-pipeline1} (d) and (e)). The exterior wall surface can be generated by lofting the newly generated B-spline curves (Figure \ref{fig:mesh-generation-pipeline1} (f)), and the smoothed discrete surface mesh can be created subsequently, following the procedures outlined above. The previously obtained luminal surface is directly adopted as the interior wall surface for the tissue.

\paragraph{Volumetric mesh generation}
Given the exterior and interior surfaces of the tissue, we may close the volume by creating planar surfaces at the inlet and outlets, which completes the B-Reps of the luminal and wall volumes. Hereafter, we refer to the planar surface on the proximal end of the artery as the \textit{inlet} and the rest planar surfaces as \textit{outlets} for the multi-branched tubular geometry. In specific, the tissue wall volume is bounded by the interior wall surface (a.k.a. the luminal surface), the exterior wall surface, and planar annular surfaces at an inlet and outlets. The luminal volume is bounded by the interior wall surface and the planar surfaces at the inlet and outlets. Mesh generation algorithms can be used to complete the volumetric mesh \cite{Antiga2008,Geuzaine2009,Si2015}. The following outlines the specific steps for the vascular FSI mesh generation.

\begin{figure}
\begin{center}
\includegraphics[angle=0, trim=0 0 0 0, clip=true, scale = 0.4]{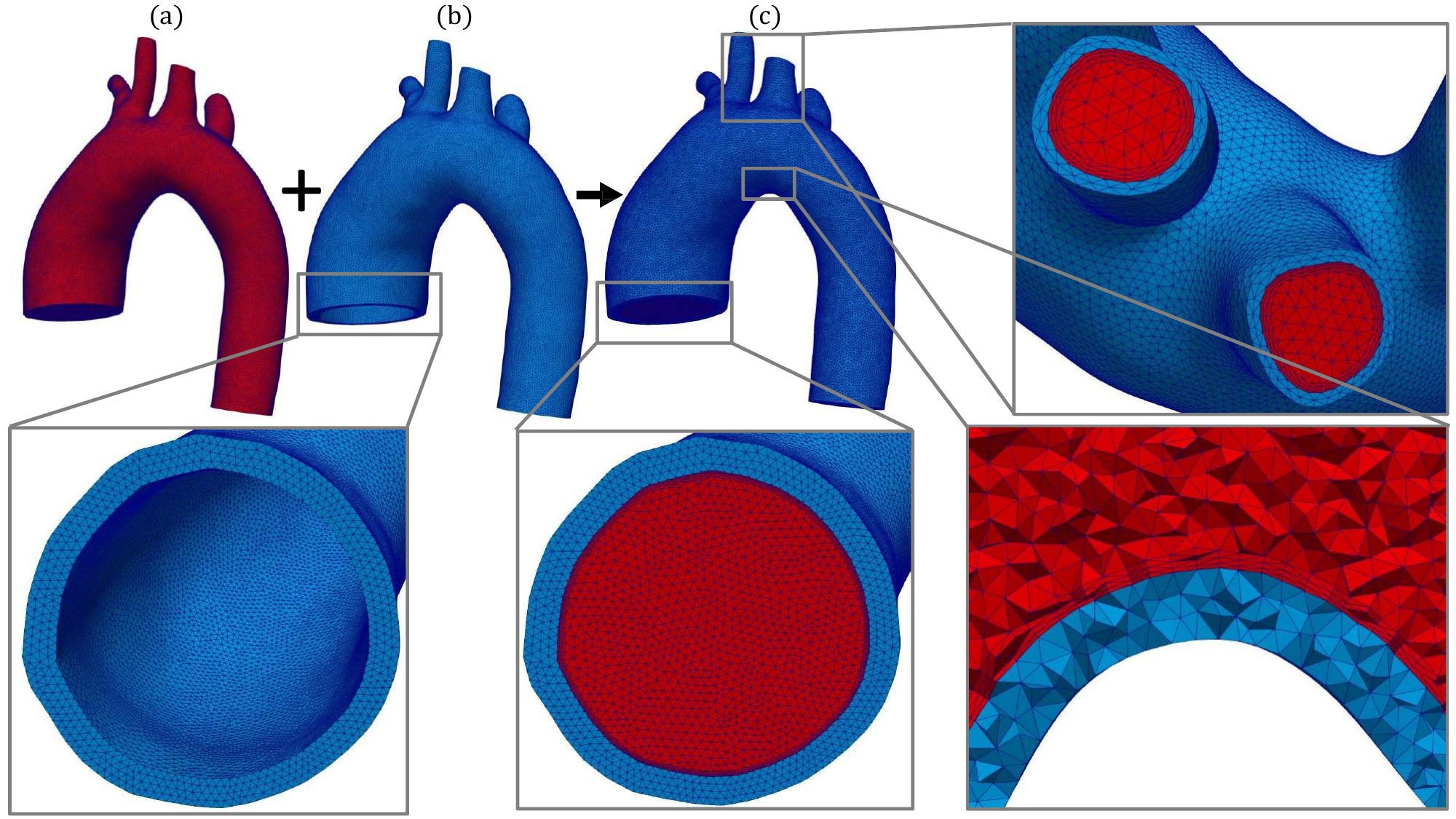}
\end{center}
\caption{(a) Generating the volumetric mesh for the lumen; (b) closing the annular planar surfaces to form the vessel wall, followed by the generation of its volumetric mesh; (c) combining the meshes to obtain the vascular FSI mesh.}
\label{fig:mesh-generation-pipeline2}
\end{figure}

\begin{enumerate}
	\item To resolve the near-wall viscous layer, a boundary resolving mesh is generated by extruding inward from the surface mesh of the lumen using the advancing layer method \cite{Garimella2000} with a specified number of layers and layer growth ratio. In particular, it is allowed to modify the luminal surface mesh to accommodate the boundary layer mesh generation in this step. This is crucial for robust BL mesh generation. The updated surface mesh will remain unchanged for the subsequent steps.
	\item Enclose the luminal volume by generating planar surface meshes at the inlet and outlets.
	\item Generate the volumetric mesh for the lumen (Figure \ref{fig:mesh-generation-pipeline2} (a)). 
	\item Enclose the tissue volume by generating the planar annular surface meshes at the inlet and outlets.
	\item Generate the volumetric mesh for the vascular tissue (Figure \ref{fig:mesh-generation-pipeline2} (b)), and ensure that the generated mesh strictly respects its B-Rep.
\end{enumerate}
The two meshes together constitute a complete mesh for FSI analysis (Figure \ref{fig:mesh-generation-pipeline2} (c)). We emphasize that mesh modification is only allowed in the first step, and the updated mesh is maintained for the subsequent steps. This is crucial for ensuring the $C^0$-continuity of the mesh across the two domains, a necessary condition for proper coupling in FSI analysis.

\begin{remark}
In contrast to mesh extrusion techniques \cite{Marchandise2013,Raut2015,Wu2022}, the above approach exhibits greater robustness and precision. It generates the CAD model of the tissue surface by manipulating the contours of each 2D segmentation, followed by mesh generation, rather than directly manipulating lumen surface meshes to create tissue meshes. This effectively prevents self-intersection for complex geometric shapes with multiple branches. Furthermore, the above strategy allows convenient adjustment of local wall thickness.
\end{remark}

\begin{remark}
The idea of segmenting interior and exterior wall surfaces has been utilized for studying the correlation between hemodynamics and plaque progression based on black-blood MRI \cite{Ladak2001,Steinman2002}. Similar modeling approaches have been combined with FSI simulations for investigating atherosclerotic plaques in coronary and carotid arteries, using multi-contrast MR \cite{Tang2004} or intravascular ultrasound images \cite{Yang2009}. Nevertheless, their emphasis was placed on the construction of the wall model with the plaques, and the geometries were relatively simple.
\end{remark}

\subsection{Definition of local basis vectors}
\label{sec:Local_basis}
\begin{figure}
\centering
\includegraphics[angle=0, trim=0 0 0 0, clip=true, scale = 0.465]{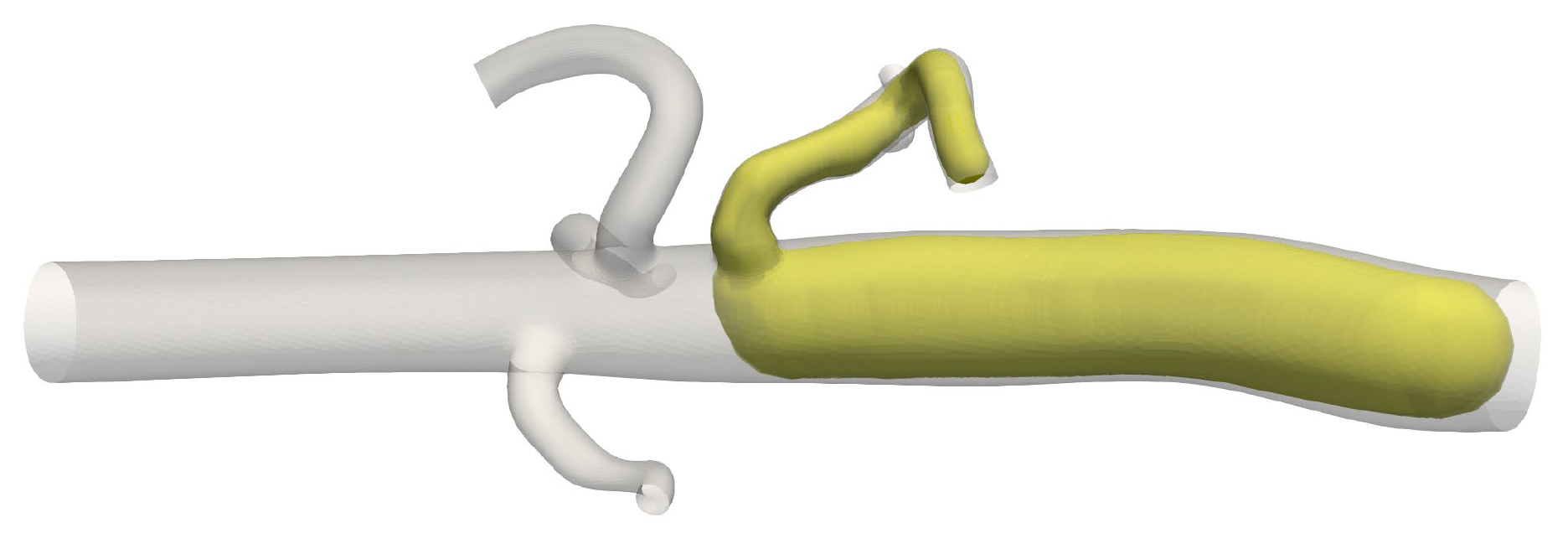}
\caption{Illustration of the canal surface associated with a centerline in the abdominal aorta.} 
\label{fig:tube}
\end{figure}

\begin{figure}
	\begin{center}
		\includegraphics[angle=0, trim=160 90 170 0, clip=true, scale = 0.95]{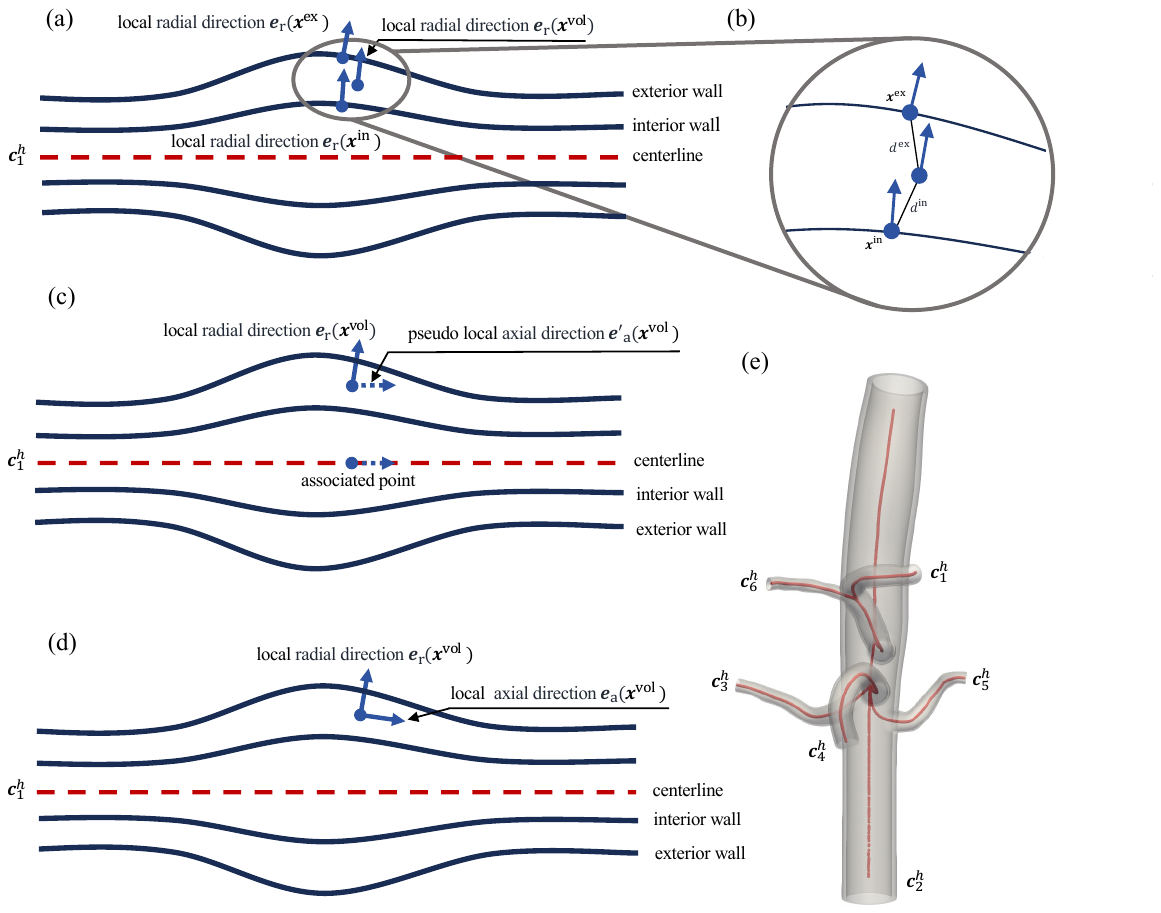} 
	\end{center}
	\caption{The generation of local (a) \& (b) radial, (c) pseudo-axial, and (d) axial vectors. (e) The red lines depict the centerlines of a multi-branched vascular model.}
	\label{fig:local basis gen}
\end{figure}

\paragraph{Vascular centerline extraction}
To facilitate the generation of local basis vectors, we need to extract the centerline of the vascular tissue. For a vascular network with $N_{\mathrm{b}}$ branches, there exists $N_{\mathrm{b}}$ centerlines that extend from the inlet to an outlet. Solving the Eikonal equation yields the centerline curve $\bm{c}_k(s)$ and the radius $r_k(s)$ of the maximum inscribed sphere within the exterior wall along this curve \cite{Piccinelli2009}. Here, $s \in [0,1]$ denotes the curvilinear abscissa along the centerline curve, and $1\leq k \leq N_{\mathrm b}$ is the branch index. Based on the above, a scalar function $\theta_{k} : \mathbb{R}^3 \rightarrow \mathbb{R}$ associated with each centerline can be defined as
\begin{align*}
\theta_{k}(\bm{x})=\min_{s\in [0,1]}\Big[\|\bm{x}-\bm{c}_k(s)\|^2-r_k^2(s) \Big],
\end{align*}
where $\bm{x}$ denotes the spatial coordinate of a generic point, and $\|\cdot\|$ represents the Euclidean distance in $\mathbb{R}^3$. The function $\theta_{k}$ is referred to as the tube function, and the set $\{\bm x : \theta_{k}(\bm x)=0\}$ gives the canal surface, which is the envelope of the inscribing spheres (Figure \ref{fig:tube}). The tube function and canal surface will assist us in generating the local basis vectors. In practice, each centerline can be obtained through the numerical approximation of the Eikonal equation, resulting in a discrete representation $\bm c^h_k(s)$ composed of a series of continuous line segments. Each segment is determined by neighboring points, and the discrete centerline $\bm c^h_k(s)$ can be represented by the centerline point set $\{\bm{x}_i^{k}\}_{i=1}^{N_{k}}$. Each point $\bm x_i^k$ in this set is also associated with a positive scalar $r_i^{k}$ representing the radius of the maximum inscribed sphere within the exterior wall surface centered at the point. All centerlines from different branches are collectively denoted as $\{\bm c^h_k(s)\}_{k=1}^{N_{\mathrm{b}}}$ (Figure \ref{fig:local basis gen} (e)).

\paragraph{Generation of local basis vectors for a single vessel}

The Frenet frame of the centerline does not always accurately reflect the local intrinsic geometry of the arterial wall. For example, in an artery with a fusiform aneurysm, the local axial direction within the aneurysm can significantly deviate from the trajectory of the centerline. To address this issue, we propose the following procedure for generating local basis vectors of a single vessel with its centerline $\bm c_{1}^h(s)$ through the following three steps.

First, for a mesh node on the interior or exterior wall surfaces, the local radial direction can be defined by the normal vector at that node, pointing towards the exterior side of the wall. This is achieved by algorithms developed for polygon surfaces \cite{Botsch2010,Jin2005} or by solving a differential equation on the surface \cite{Cenanovic2020}. For a generic point $\bm{x}^\mathrm{vol}$ located inside of the vessel wall, we first traverse all mesh nodes on the interior and exterior wall surfaces to identify the closest nodes to $\bm{x}^\mathrm{vol}$, which are denoted by $\bm{x}^\mathrm{in}$ and $\bm{x}^\mathrm{out}$, respectively. The distance between $\bm x^{\mathrm{vol}}$ and $\bm{x}^\mathrm{in}$ ($\bm{x}^\mathrm{out}$) is denoted by $d^\mathrm{in}$ ($d^\mathrm{out}$). The local radial direction at the point $\bm{x}^\mathrm{vol}$ is then defined as a weighted sum as
\begin{align*}
\bm{e}_\mathrm{r}(\bm{x}^\mathrm{vol}) := \omega^\mathrm{ex} \bm{e}_\mathrm{r}(\bm{x}^\mathrm{ex}) + \omega^\mathrm{in} \bm{e}_\mathrm{r}(\bm{x}^\mathrm{in}) \quad \mbox{ with } \quad \omega^\mathrm{ex}:=\frac{d^\mathrm{in}}{d^\mathrm{ex}+d^\mathrm{in}} \quad \mbox{and} \quad \omega^\mathrm{in}:=\frac{d^\mathrm{ex}}{d^\mathrm{ex}+d^\mathrm{in}}.
\end{align*}
In the above, $\omega^\mathrm{ex}$ and $\omega^\mathrm{in}$ are the relative weights based on the distances (Figure \ref{fig:local basis gen} (a) and (b)).

Second, to obtain the local circumferential direction at the point $\bm{x}^\mathrm{vol}$, a pseudo axial direction is first introduced as follows. We traverse the centerline point set  $\{\bm{x}_i^{1}\}_{i=1}^{N_{1}}$ and find the point with the minimum distance to the point $\bm{x}^\mathrm{vol}$ on the wall, which is referred to as the \textit{associated point} of $\bm{x}^{\mathrm{vol}}$. This search procedure is abstractly represented by a function $\mathcal {F}(\bm{x}^\mathrm{vol}, \{\bm{x}_i^{1}\}_{i=1}^{N_{1}})$ that returns the coordinate of the associated point. The pseudo local axial direction $\bm{e}'_{\mathrm{a}}(\bm{x}^\mathrm{vol})$ is defined by the tangential vector at the associated point in the centerline, and it points in the flow direction (i.e., from the inlet to the outlet). We mention that this vector is not necessarily perpendicular to the local radial direction but may assist defining the local circumferential direction $\bm{e}_c(\bm{x}^\mathrm{vol})$ by the cross product $\bm{e}'_{\mathrm{a}}(\bm{x}^\mathrm{vol}) \times \bm{e}_\mathrm{r}(\bm{x}^\mathrm{vol})$ (Figure \ref{fig:local basis gen} (c)). 

Third, given the local radial direction $\bm{e}_\mathrm{r}(\bm{x}^\mathrm{vol})$ and local circumferential direction $\bm{e}_\mathrm{c}(\bm{x}^\mathrm{vol})$ at the point $\bm{x}^\mathrm{vol}$, the local axial direction $\bm{e}_\mathrm{a}(\bm{x}^\mathrm{vol})$ is eventually defined by their cross product $\bm{e}_\mathrm{r}(\bm{x}^\mathrm{vol}) \times \bm{e}_\mathrm{c}(\bm{x}^\mathrm{vol})$ (Figure \ref{fig:local basis gen} (d)).

\begin{figure}
\begin{center}	
\includegraphics[angle=0, trim=90 110 110 100, clip=true, scale = 0.7]{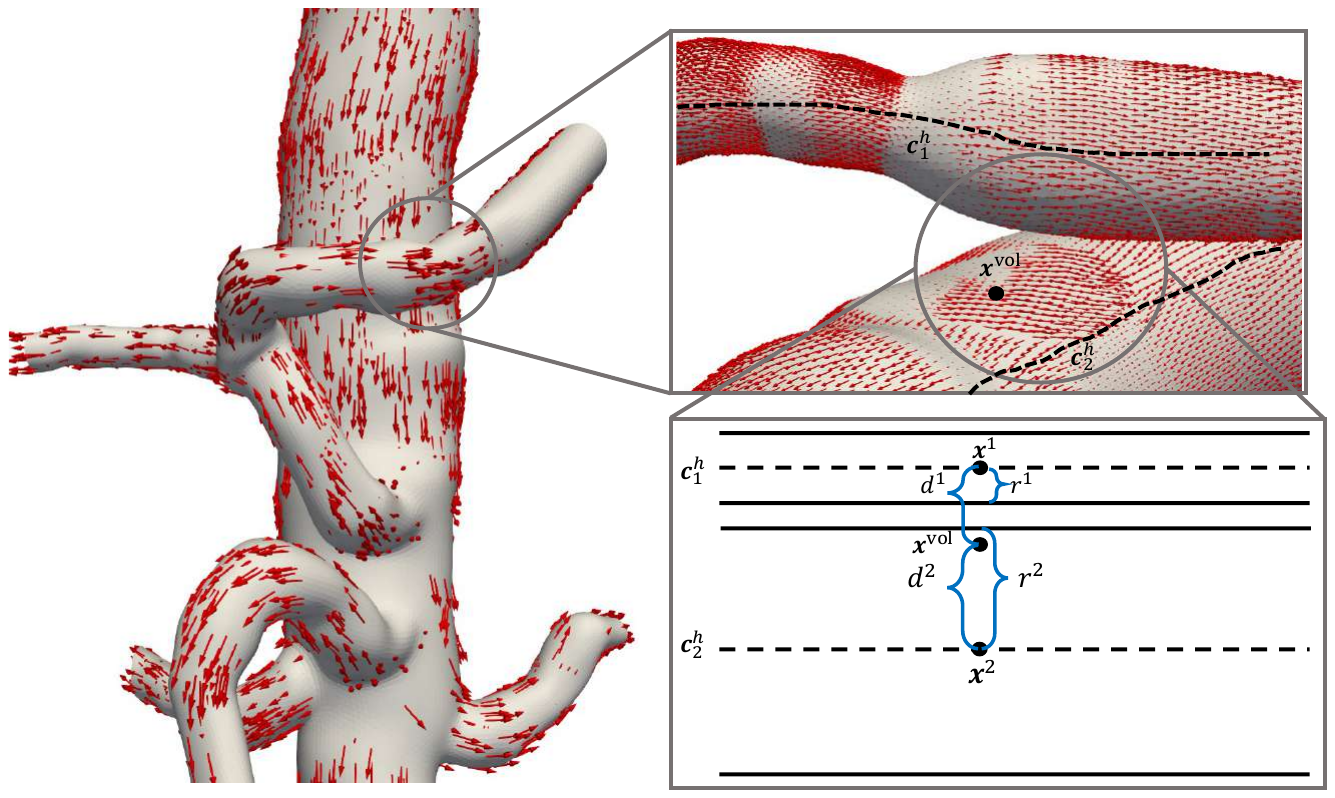}
\end{center}
\caption{The misidentification of centerline using the absolute distance for two non-intersecting branches, with centerlines $\bm{c}_{2}^h$ and $\bm{c}_{6}^h$, being very close to each other.}
\label{fig:challenge1}
\end{figure}

\paragraph{Generation of local basis vectors for a multi-branched vessel}
When dealing with a single vessel, the previously mentioned strategy is adequate. Yet, for a vascular network with $N_{\mathrm{b}}$ centerlines, a point with the minimum distance can be found on each centerline. There is thus a need to define a unique associated point from these $N_{\mathrm{b}}$ points. A straightforward method is based on the absolute distance, that is, to identify the point with the closest distance to point $\bm{x}^{\mathrm{vol}}$ among these $N_{\mathrm{b}}$ points as the associated point. Nevertheless, when two branches are close to each other, such as the branches with centerlines $\bm{c}_{1}^h(s)$ and $\bm{c}_{2}^h(s)$ depicted in Figure \ref{fig:challenge1}, the closest point $\bm{x}^{1}$ on the centerline $\bm{c}^h_1(s)$ is identified as the associated point when $d^{1} < d^{2}$, which results in the generation of non-physiological local basis vectors. To address this issue, we propose a refined metric that utilizes the local radius information recorded on the centerline points. Specifically, for each of the $N_{\mathrm{b}}$ points, the relative distance ${d^{k}}/{r^{k}}$ to the point $\bm{x}^{\mathrm{vol}}$ is calculated. The value of $r^{k}$ is the radius at the closest point on the $k$-th centerline, and it is at a distance of $d^{k}$ from the point $\bm{x}^{\mathrm{vol}}$. The relative distance will be significantly greater than $1.0$ for points on unrelated centerlines. Consequently, the point with the smallest relative distance is recognized as the associated point. The implementation of this procedure is detailed in Algorithm \ref{alg:Correct_Associated_Point}. 

\renewcommand{\algorithmicrequire}{\textbf{Input:}}
\renewcommand{\algorithmicensure}{\textbf{Output:}}
\begin{algorithm}
	\caption{Associated point identification for multiple centerlines}
	\label{alg:Correct_Associated_Point}
	\begin{algorithmic}[1]
		\Require
		\Statex A generic point $\bm{x}^\mathrm{vol}$ located inside of the vessel wall
		\Statex All discrete centerlines with their centerline point sets $\{\bm{x}_i^{k}\}_{i=1}^{N_{k}}$ and radius sets $\{r_i^{k}\}_{i=1}^{N_{k}}$
		\Ensure
		\Statex Associated point of $\bm{x}^{\mathrm{vol}}$
		\State Initialize $min\_ratio \gets \mathrm{INT\_MAX}$
		\State Initialize $associated\_point \gets \mathrm{NULL}$
		\For{$k \gets 1$ to $N_{\mathrm{b}}$}
		\State $\bm{x}_\mathrm{closest}^{k} \gets \mathcal{F} ( \bm{x}^{\mathrm{vol}}, \{\bm{x}_i^{k}\}_{i=1}^{N_{k}} )$
		\State Absolute distance $d^{k} \gets \|\bm{x}^{\mathrm{vol}} - \bm{x}_\mathrm{closest}^{k}\|$
		\State Relative distance $ratio \gets d^{k}/r_\mathrm{closest}^{k}$
		\If{$ratio < min\_ratio$}
		\State $min\_ratio \gets ratio$
		\State $associated\_point \gets \bm{x}_\mathrm{closest}^{k}$
		\EndIf
		\EndFor
	\end{algorithmic}
\end{algorithm}

\begin{remark}
The inherent challenge in generating local basis vectors for multi-branched vessels based on centerlines lies in identifying the correct centerline. An alternative approach was presented in \cite{Roy2014}, in which a cone defined by a prescribed angle was utilized to select the correct centerline.
\end{remark}

\begin{figure}
\centering
\includegraphics[angle=0, trim=0 0 320 0, clip=true, scale = 0.5]{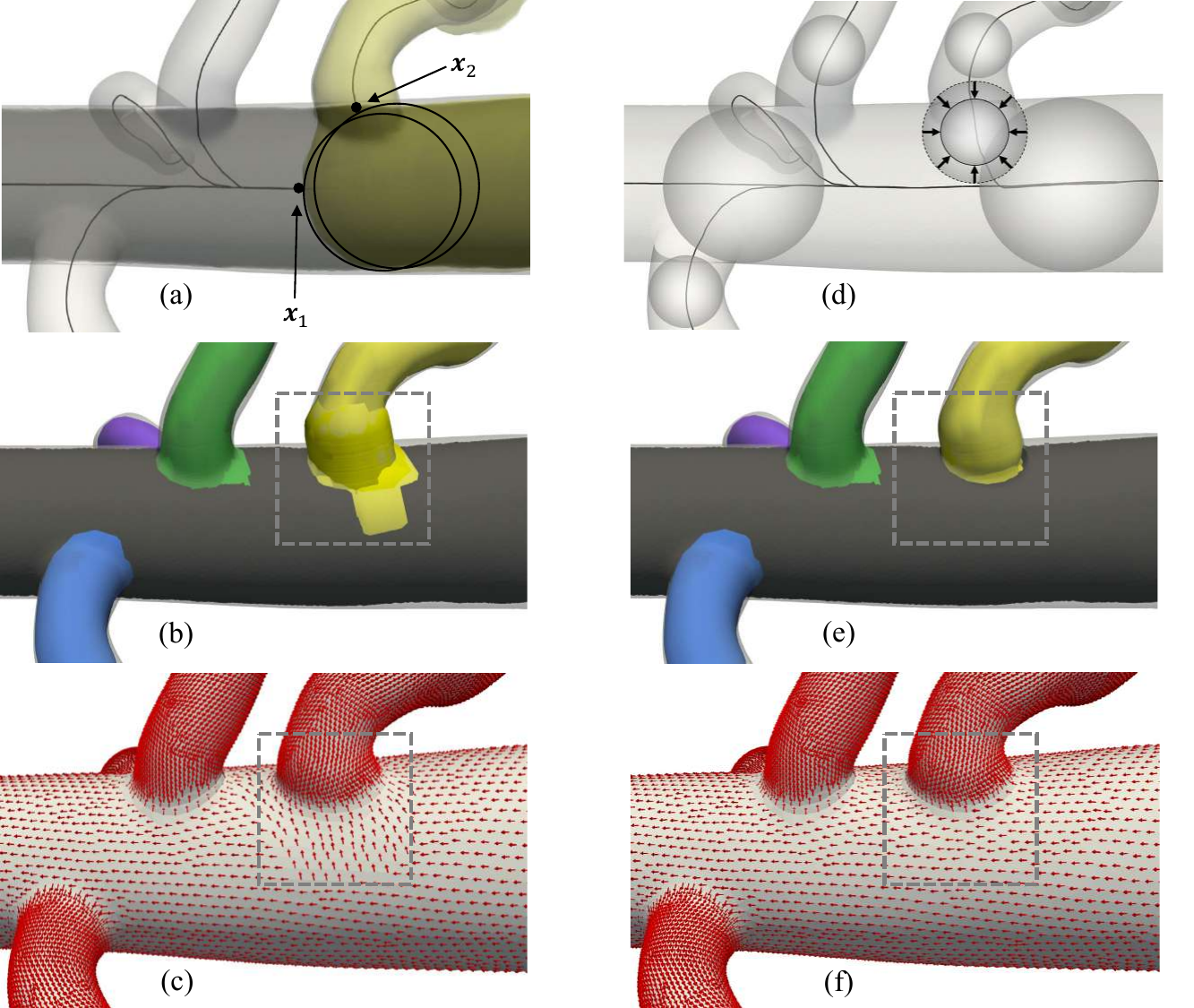}
\caption{The bifurcation region requires further refinement of local radii of child branches.}
\label{fig:challenge2}
\end{figure}

\paragraph{Treatment at the bifurcation region}
Near bifurcation regions when one branch is significantly larger than the other, using the relative distance leads to a tendency to misidentify the associated point. We elaborate on this situation and propose a refined strategy to address it in the following.

The bifurcation region is defined as the location where one centerline intersects the canal surface of another centerline. Consider two centerlines $\bm c^h_1(s)$ and $\bm c^h_2(s)$ that constitute a bifurcation region. We may identify one as the \textit{parent} and the other one as the \textit{child} branch according to their local radii near the bifurcation. Let the intersection point between the canal surface of $\bm c^h_1(s)$ and the centerline $\bm c^h_2(s)$ be $\bm x_1$. Along the centerline $\bm c^h_1(s)$, there exists a sphere that touches the point $\bm x_1$ due to the definition of the canal surface, and we denote the radius of the sphere as $r_1$. Conversely, the intersection point between the canal surface of $\bm c^h_2(s)$ and the centerline $\bm c^h_1(s)$ is denoted by $\bm x_2$, and the radius of the corresponding sphere is denoted as $r_2$. With these local geometric information, we may identify $\bm c^h_1(s)$ and $\bm c^h_2(s)$ as the parent and child branches if $r_1 > r_2$; otherwise, if $r_1 < r_2$, $\bm c^h_1(s)$ is the child branch, and $\bm c^h_2(s)$ is the parent branch.

Due to the nature of the Eikonal Equation, the local radius of the child branch smoothly transits from large to small values near the bifurcation region. This causes a non-physiological overestimation of the local radius, as shown in Figure \ref{fig:challenge2} (a). This issue is visualized by inspecting the canal surface of different centerlines shown in Figure \ref{fig:challenge2} (b). As a result, there is a tendency to locate the associated point on the child branch near the bifurcation when using the relative distance, causing incorrect local basis vector assignment on the parent branch, as illustrated in Figure \ref{fig:challenge2} (c). 

To resolve this issue, we first identify if the child centerline point is located within the canal surface of the parent centerline. This can be achieved by evaluating the relative distance of the point to the parent centerline. When these points are identified (i.e. points with the relative distance less than $1.0$), their recorded radius is reduced by a scaling function $\phi$, which is a univariate function of the relative distance. The function $\phi$ is a smooth, monotonically increasing function and satisfies $\phi(1) = 1$ and $\phi(0) = \varepsilon$. Here, $\varepsilon$ is a small positive number, ensuring $\phi(0)$ is not zero to avoid division-by-zero errors in the subsequent calculation of the relative distance. The specific criteria for identifying the bifurcation region and the algorithm for reducing the radii from the child branches are detailed in Algorithm \ref{alg:Update Radii}. As shown in Figure \ref{fig:challenge2} (d), the local radii recorded on the child branch at the bifurcation are reduced, and the refined tube surfaces also indicate that the erroneous influence of the child branch has been corrected (Figure \ref{fig:challenge2} (e)). Employing the relative distance with the radius values modified, the correct local basis vectors are generated (Figure \ref{fig:challenge2} (f)). We mention that for a set of multiple centerlines, these procedures have to be performed for each bifurcation region.

\begin{figure}
	\begin{center}
		\includegraphics[angle=0, trim=50 100 30 90, clip=true, scale = 0.5]{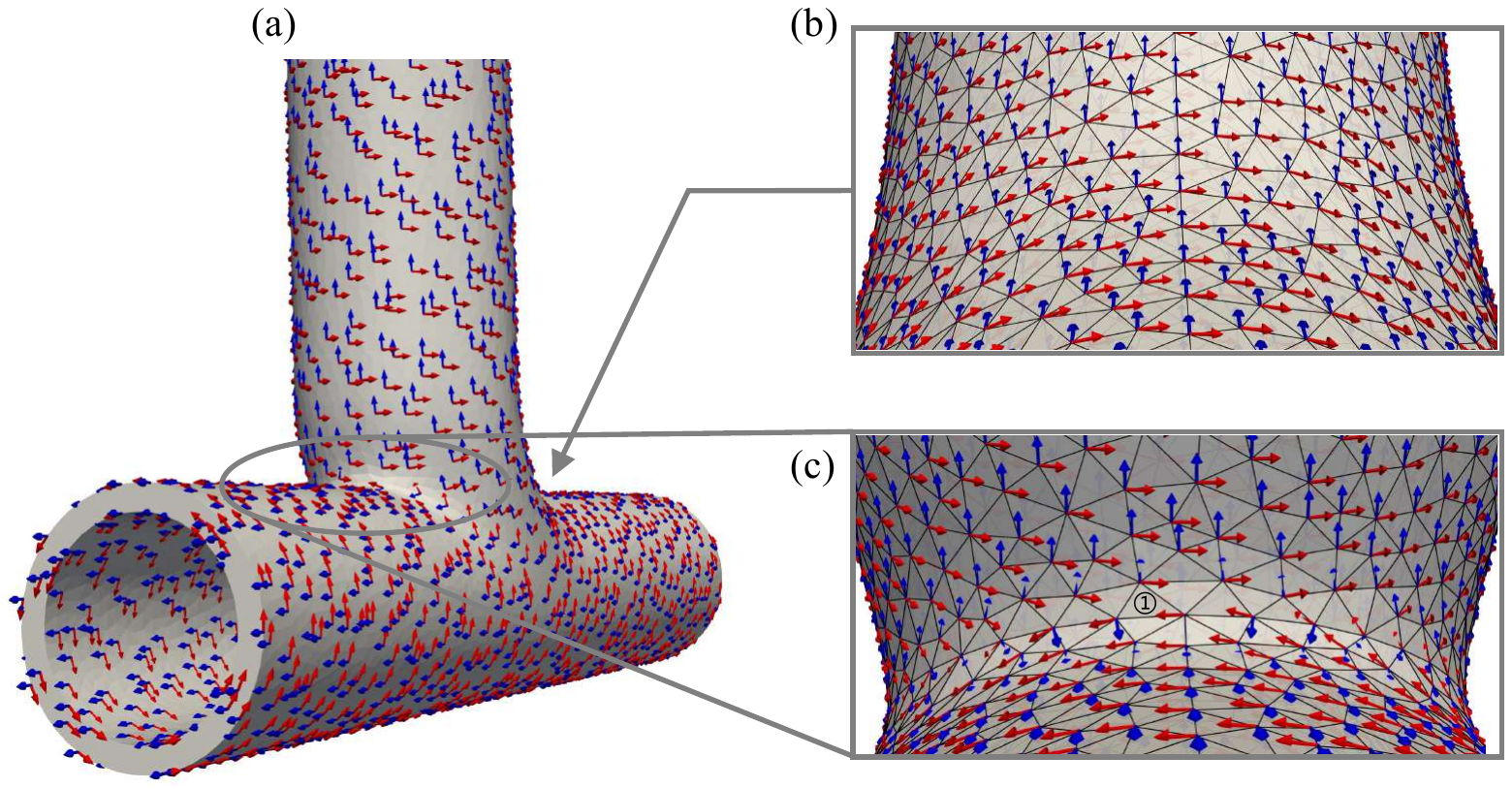}
	\end{center}
	\caption{The local basis vectors defined on the mesh node. The red and blue arrow represent the local circumferential and the local axial basis vector, respectively.}
	\label{fig:local_basis_node}
\end{figure}

\begin{algorithm}
	\caption{Modification of the radii of one child branch centerline points}
	\label{alg:Update Radii}
	\begin{algorithmic}[1]
		\Require 
		\Statex Parent branch centerline with its centerline point set $\{\bm{x}_i^{\alpha}\}_{i=1}^{N_{\alpha}}$ and radius set $\{r_i^{\alpha}\}_{i=1}^{N_{\alpha}}$
		
		\Statex Child branch centerline with its centerline point set $\{\bm{x}_i^{\beta}\}_{i=1}^{N_{\beta}}$ and radius set $\{r_i^{\beta}\}_{i=1}^{N_{\beta}}$
		
		\Statex Scaling function $\phi$
		\Ensure 
		\Statex Updated radii for child branch centerline points
		\For{$j \gets 1$ to $N_{\beta}$}
		\State $\bm{x}_\mathrm{closest}^{\alpha} \gets \mathcal{F}$($\bm{x}_j^{\beta}, \{\bm{x}_i^{\alpha}\}_{i=1}^{N_{\alpha}}$)
		\State $d_j^{\beta} \gets \| \bm{x}_j^{\beta} - \bm{x}_\mathrm{closest}^{\alpha}\|$
		\If{$d_j^{\beta}/r_{\text{closest}}^{\alpha} < 1.0$}
		\State $r_j^{\beta} \gets r_j^{\beta} \phi(d_j^{\beta}/r_{\text{closest}}^{\alpha}) $
		\EndIf
		\EndFor
	\end{algorithmic}
\end{algorithm}

\begin{remark}
It is possible to have two pairs of local basis vectors pointing in opposite directions at the mesh nodes of an element near the bifurcation region (e.g., see the element \ding{192} in Figure \ref{fig:local_basis_node}). In such a case, if we calculate the local circumferential and axial basis vectors at the quadrature points through interpolation, the definition of the local basis vectors will become singular. In our practice, the local basis vectors are generated at the quadrature points directly.	
\end{remark}

\subsection{Constitutive model for arterial wall}
\label{sec:constitutive-model}
In this section, we introduce the arterial wall constitutive models. The arterial wall is modeled as a slightly compressible or fully incompressible hyperelastic material. Given the referential configuration $\Omega_0 \subset \mathbb R^3$, its motion at time $t$ is given by an orientation-preserving one-to-one mapping $\bm \varphi_t(\cdot)$ that maps the body to the current configuration $\Omega_t = \bm \varphi_t(\Omega_0)$. The kinematics is locally described by the deformation gradient $\bm F := \partial \bm \varphi_t / \partial \bm X$ for $\bm X \in \Omega_0$. Its determinant is known as Jacobian and denoted by $J$. The right Cauchy-Green deformation tensor is given by $\bm C:= \bm F^{T}\bm F$. On the referential configuration, a scalar field $P$ is introduced to represent the thermodynamic pressure. The material behavior is characterized by the free energy of the form, 
\begin{align}
\label{eq:G_split}
G(\bm C, P) = G_{\mathrm d}(\bm C) + G_{\mathrm v}(P),
\end{align}
where $G$ represents the Gibbs free energy. It is a scalar-valued function and is additively decomposed into the contributions of $G_{\mathrm{d}}$ and $G_{\mathrm{v}}$. The deformation is often decoupled into isochoric and volumetric parts following the decomposition of the Flory type, i.e., $\bm F = J^{1/3}\tilde{\bm F}$, and correspondingly the unimodular right Cauchy-Green deformation tensor is defined as $\tilde{\bm C} := \tilde{\bm F}^T \tilde{\bm F}$. If the energy function $G_{\mathrm a}$ depends on $\tilde{\bm C}$, it can be interpreted as the energy associated with isochoric deformation, and $G_{\mathrm b}$ is then the bulk or volumetric free energy. Through a Coleman-Noll-type analysis \cite{Liu2018}, we may obtain the constitutive relations for the second Piola-Kirchhoff stress $\bm S$, the Cauchy stress $\bm \sigma$, and the density $\rho$ based on the free energy $G$ as
\begin{align}
\bm S = 2 \frac{\partial G_{\mathrm d}}{\partial \bm C} - PJ \bm C^{-1}, \quad \bm \sigma = \bm \sigma_{\mathrm d} - p\bm I, \quad \bm \sigma_{\mathrm d} = 2 J^{-1} \bm F \frac{\partial G_{\mathrm d}}{\partial \bm C} \bm F^T,   \quad \mbox{and} \quad \rho = \rho_0 ( \frac{\partial G_{\mathrm v}}{\partial P})^{-1}.
\end{align}
In the theory constructed based on the Gibbs free energy, the pressure $P$ acts as a primitive variable, rendering a mixed formulation suitable for (nearly-)incompressible materials \cite{Liu2018}. If the energy is taken simply as $G_{\mathrm b} = P$, the one has $\rho = \rho_0$ and the material is fully incompressible. In this work, we adopt the following form for the volumetric energy,
\begin{align}
\label{eq:G_v}
 G_{\mathrm{v}}(P) = \frac{-P^2+P\sqrt{P^2+K^2}}{2K\rho_0}-\frac{K}{2\rho_0} \ln \left(\frac{\sqrt{P^2+K^2}-P}{K}\right), \quad   \rho = \frac{\rho_0}{K}\left(\sqrt{P^2+K^2}+P\right),
\end{align} 
where $K$ is the bulk modulus. This volumetric energy is related to the volumetric energy $K(J^2-1-2\ln(J))/4$ through a Legendre transformation \cite{Liu2018}. A specific functional form of $G_{\mathrm d}$ completes the definition of the constitutive model.

\paragraph{Neo-Hookean model} It is a common choice to model the vascular wall as an isotropic hyperelastic material, and the neo-Hookean model is given by
\begin{align}
\label{eq:NH constitutive model}
G_{\mathrm d}(\bm C) =  \frac{\mu}{2} \left( \tilde{I}_1 - 3 \right),
\end{align}
where $\tilde{I}_1 := \mathrm{tr}\tilde{\bm C}$ is the first principal invariant of $\tilde{\bm C}$, and the parameter $\mu$ is the shear modulus. This model enjoys a compact energy form with a single material parameter and can be rigorously derived based on the microscopic structure using the Gaussian distribution. Despite its wide usage in vascular FSI analysis \cite{Baeumler2020,Liu2020,Suito2014}, we need to point out that it was originally designed for vulcanized rubber and its predictive capability is limited to moderate deformation states \cite{Xiang2020}.

\begin{figure}
\begin{center}
\includegraphics[angle=0, trim=150 80 130 70, clip=true, scale = 0.4]{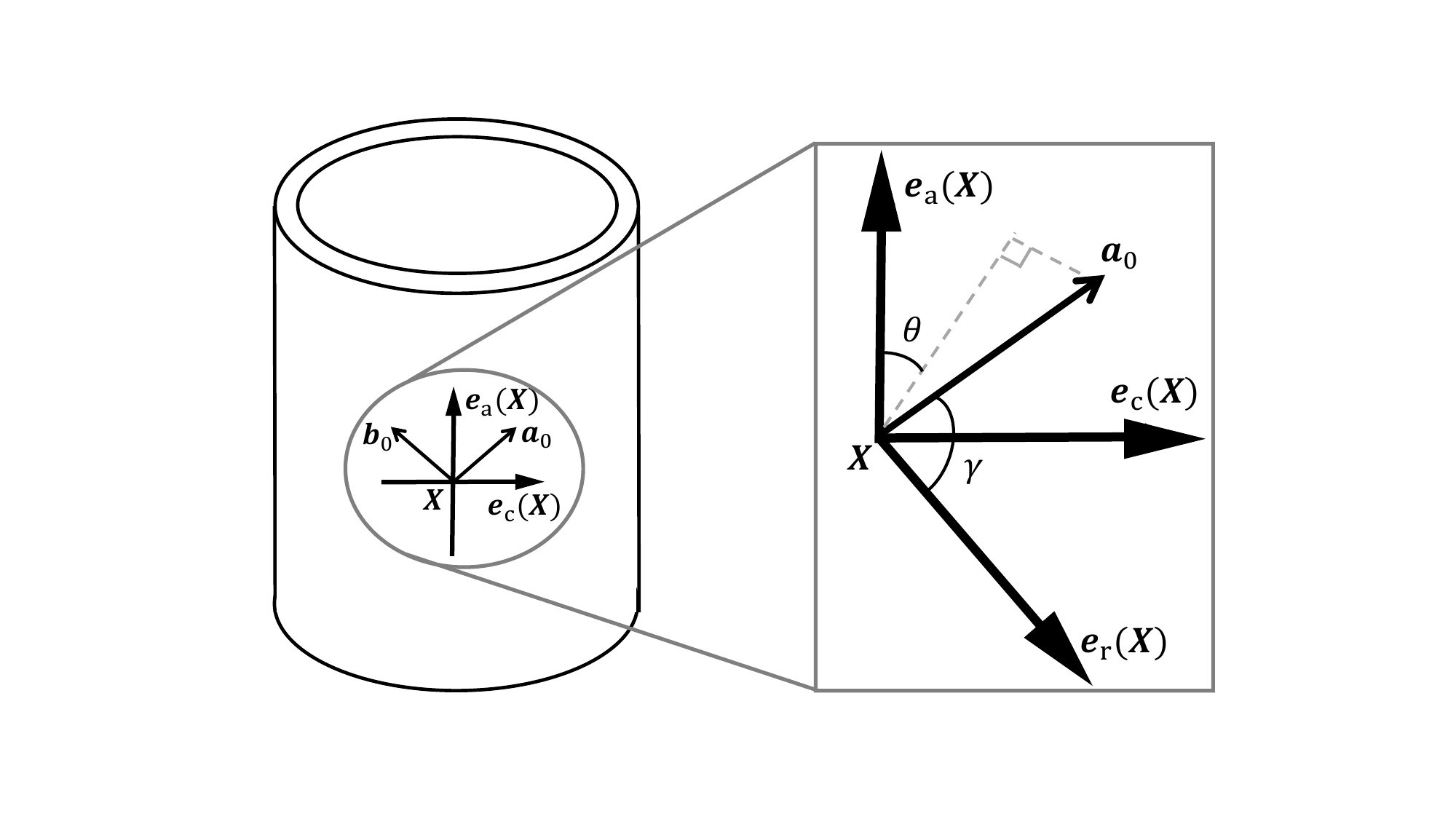}
\end{center}
\caption{Fiber orientation vectors $\bm a_{0}$ and $\bm b_{0}$ defined by local basis vectors.}
\label{fig:local_basis_diagram}
\end{figure}

\paragraph{Gasser-Ogden-Holzapfel model}
Based on histology, the artery can be modeled as a composite of ground matrix and fibers, and the potential $G_{\mathrm d}$ gets decomposed into the contribution of non-collagenous groundmatrix and two collagen fibers,
\begin{align}
\label{eq:HGO-C constitutive model}
G_{\mathrm d}(\bm C) = \frac{\mu}{2}(\tilde{I}_1-3)+\sum_{i=4,6}\frac{k_{1}}{2k_{2}}\left[\exp\left\{k_{2}\big[\kappa{I}_{1}+(1-3\kappa){I}_{i}-1\big]^{2}\right\}-1\right].
\end{align}
The above model is referred to as the Gasser-Ogden-Holzapfel (GOH) model \cite{Gasser2006}, wherein the invariants are defined as
\begin{align*}
I_1 := \mathrm{tr}\bm C, \quad I_4 := \bm C : \left( \bm a_{0}\otimes \bm a_{0}\right), \quad \mbox{and} \quad I_6 := \bm C : \left( \bm b_{0}\otimes \bm b_{0}\right),
\end{align*}
and the latter two describe the stretch along the fiber directions. The parameters $\mu$ and $k_1$  are stress‐like, representing the stiffness of the ground matrix and collagen fibers; $k_2$ is a dimensionless parameter; the unit vectors $\bm a_0$ and $\bm b_0$ represent the mean orientation of the embedded fibers; the parameter $\kappa \in [0, 1/3]$ characterizes the fiber dispersion. The first term in \eqref{eq:HGO-C constitutive model} represents the potential energy due to the ground matrix, which is treated as an isotropic neo-Hookean material; the exponential terms in \eqref{eq:HGO-C constitutive model} represent the anisotropic free energy contributions due to the two families of fibers. In this work, we consider the two families of fibers disposed in a symmetric manner with respect to the axial direction. Therefore, the fiber directions $\mathbf{a}_{0}(\bm X)$ and $\mathbf{b}_{0}(\bm X)$ at a material point $\bm X \in \Omega_0$ can be defined by the local basis vector $\{\bm{e}_{\mathrm r}(\bm X),  \bm{e}_{\mathrm c}(\bm{X}),  \bm{e}_{\mathrm a}(\bm{X}) \}$ as follow,
\begin{align*}
\bm {a}_{0}(\bm X) =&  \cos \theta \sin \gamma \bm{e}_{\mathrm a}(\bm X) +  \sin \theta \sin \gamma \bm {e}_{\mathrm c}(\bm X) +  \cos \gamma \bm{e}_{\mathrm r}(\bm X), \displaybreak[2] \\ 
\bm{b}_{0}(\bm X) =& \cos \theta \sin \gamma \bm{e}_{\mathrm a}(\bm X) - \sin \theta \sin \gamma \bm e_{\mathrm c}(\bm X) +  \cos \gamma \bm{e}_{\mathrm r}(\bm X),
\end{align*}
where $\theta \in \left[0, \pi \right]$ and  $\gamma \in \left[0, 2\pi \right]$ are the Eulerian angles, as shown in Figure \ref{fig:local_basis_diagram}.

\begin{remark}
In their original proposal, the artery model was constructed based on the fully incompressible assumption, and the potential energy adopts $\tilde{I}_4 := J^{-2/3}I_4$ and $\tilde{I}_6 := J^{-2/3}I_6$ instead of $I_4$ and $I_6$ in the definition \cite{Gasser2006}. However, that model was plagued by pathological material behavior when being extended to compressible regimes. A modification was introduced in \cite{Nolan2014} to rectify the issue and the rectified energy is given by \eqref{eq:HGO-C constitutive model}.
\end{remark}

\begin{remark}
The dispersion parameter $\kappa$ characterizes the distribution of the fibers. When $\kappa = 0$, there is no dispersion, and the model \eqref{eq:HGO-C constitutive model} recovers its predecessor \cite{Holzapfel2000}, which is known as the HGO model. When $\kappa = 1/3$, the fibers distribute equally in all directions and the model becomes isotropic and takes the exponential form, rendering it similar to the one of Demiray \cite{Demiray1972}.
\end{remark}

\subsection{Tissue prestressing and an initialization procedure}
\label{subsec:prestressing}
In patient-specific modeling, the vascular wall is imaged in a state under physiological loading, meaning that there is a stress field in the imaged configuration. The stress field is often referred to as \textit{prestress} and denoted by $\bm \sigma_{0}$. This prestress balances the in vivo blood pressure and WSS. We incorporate the prestress term $\bm \sigma_0$ into the Cauchy stress to furnish a modified constitutive relation  $\bm \sigma = \bm \sigma_{\mathrm d} -p \bm I + \bm \sigma_0$. To determine the value of $\bm \sigma_0$ at each quadrature point, we perform a fixed-point iteration so that the wall does not exhibit deformation under the loading of the fluid. A rigid-walled simulation is first conducted to acquire the physiological pressure and WSS distributions. We mention that the patient-specific image data investigated in this work is indeed cardiac-gated and synchronized to the cardiac cycle, which guarantees the compatibility of the geometric model of the wall and the prestress to be determined. Upon attaining a steady-state flow state, the solution is loaded as the input data to generate the prestress $\bm \sigma_0$. The algorithm is detailed below, drawing inspiration from a prior work grounded in pure displacement formulation \cite{Hsu2011}. 

\begin{algorithm}
	\caption{Prestress generation}
	\label{alg:Ps_Gen}
	\begin{algorithmic}[1]
		\Require 
		\Statex The diastolic physiological pressure and WSS distribution obtained through the rigid wall simulation
		\Statex A prescribed tolerance $\mathrm{tol}_{\mathrm{P}}$
		\Statex Maximum number of iterations $m_{\mathrm{max}}$
		\Ensure 
		\Statex The prestress $\bm \sigma_0$
		\State Initialize $\bm \sigma_{0, (0)} \gets  \bm 0$
		\For{$m \gets  0$ to $m_{\mathrm{max}}$}
		\State $\bm \sigma_0 \gets  \bm \sigma_{0, (m)}$
		\State From $t_m$ to $t_{m+1}$, solve the FSI problem using the backward Euler method for temporal integration
		\State Update the prestress tensor as $\bm \sigma_{0, (m+1)} \gets  \bm \sigma_{\mathrm{d}}(\bm u_{m+1}) - p_{m+1}\bm I + \bm \sigma_{0, (m)}$
		\If{$\| \bm u_{m+1} \|_{\mathfrak l_{\infty}} \leq \mathrm{tol}_{\mathrm{P}}$}
		\State $\bm \sigma_0 \gets \bm \sigma_{0,(m+1)}$
		\State \textbf{break}
		\EndIf
		\EndFor
	\end{algorithmic}
\end{algorithm}

\begin{remark}
Alternative methods exist for determining prestress. For instance, one approach suggests establishing the zero-pressure configuration initially, which subsequently deforms to the imaged configuration under physiological loading \cite{Takizawa2010a}. Another approach calculates a hypothetical deformation gradient to determine prestress \cite{Gee2010}. We adopt here the approach that directly determines the prestress field, as it is widely acknowledged that prestress originates from the tissue growth and remodeling rather than elastic deformation \cite{Humphrey2002}.
\end{remark}

\section{Results}
We begin by evaluating the local basis vectors generated using the method presented in Section \ref{sec:Local_basis} to demonstrate the effectiveness of the proposed morphology-based strategy. After that, two vascular FSI studies are considered, focusing on the impact of the fiber-reinforced model on the prediction of biomechanical factors. The key features of our FSI model can be summarized as follows.
\begin{enumerate}
	\item The formulation is based on the unified continuum formulation, allowing a convenient way of monolithic coupling \cite{Liu2018,Sun2024}. The mesh motion in the fluid subdomain is governed by the harmonic extension algorithm.
	\item The blood is modeled as a Newtonian fluid.
	\item The variational multiscale formulation is employed as the numerical model to provide pressure stabilization for both fluids and solids and to facilitate large-eddy simulation of complex blood flows  \cite{Liu2018,Liu2020}.
	\item Linear tetrahedral element is used for the interpolation of all field variables.
	\item A JWH-generalized-$\alpha$ method is used for time integration \cite{Sun2024}. The spectral radius of the amplification matrix at the highest mode $\varrho_{\infty}$ is set to $0.0$ to ensure robust time integration.
\end{enumerate}

\begin{figure}
\begin{center}
\begin{tabular}{cc}
\includegraphics[angle=0, trim=20 10 50 90, clip=true, scale = 0.26]{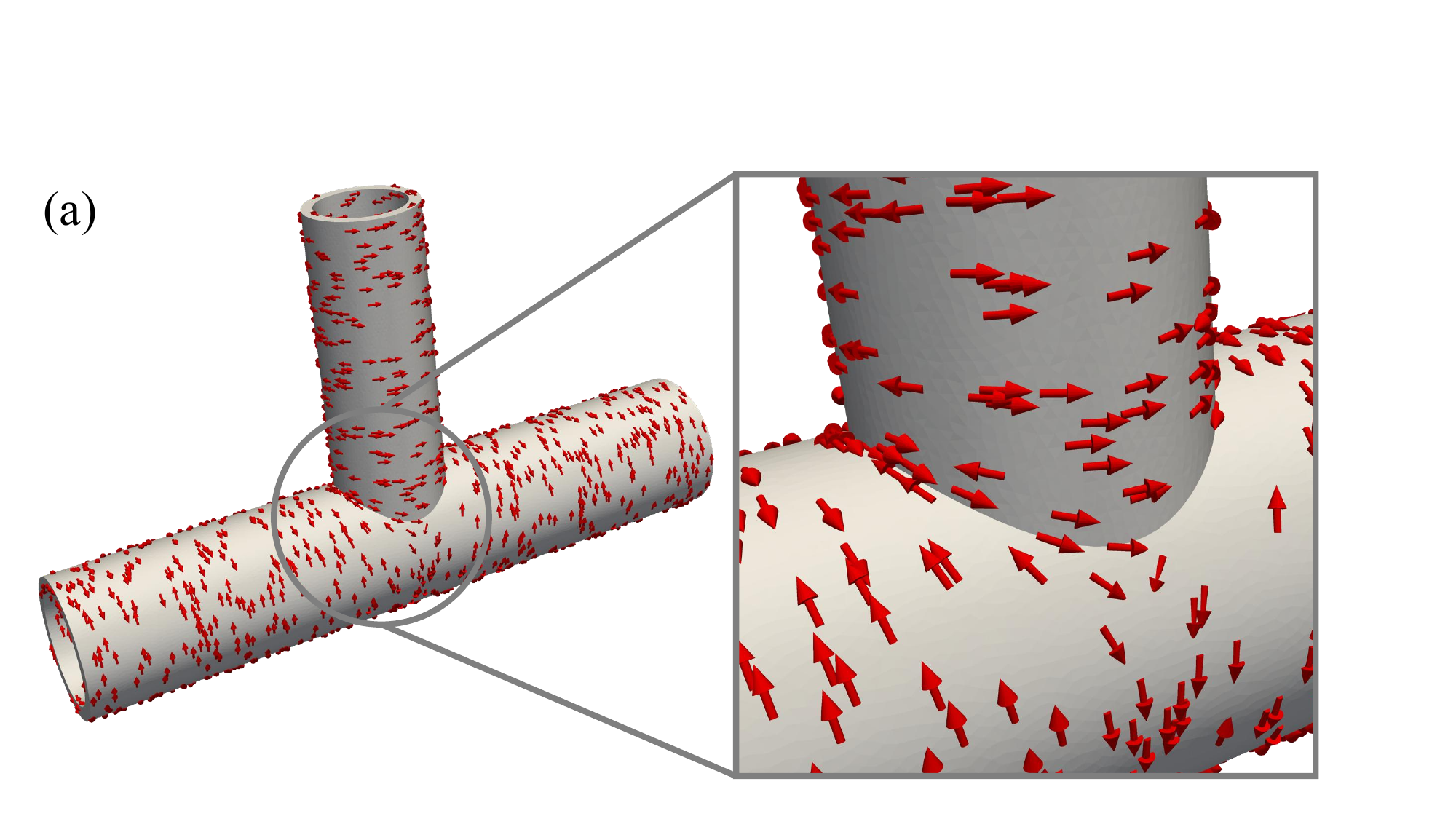} & \includegraphics[angle=0, trim=20 10 50 90, clip=true, scale = 0.26]{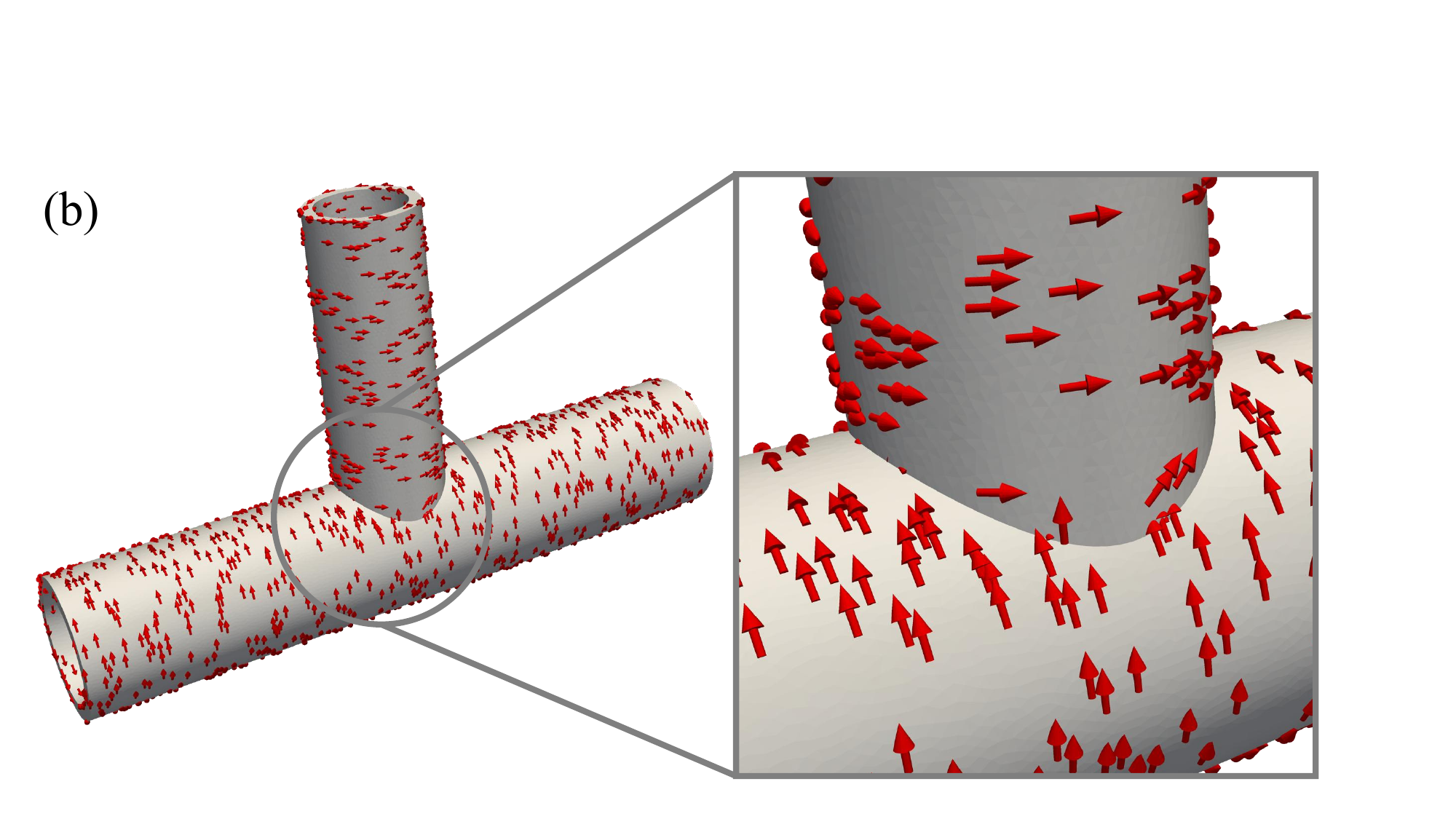} \\
\includegraphics[angle=0, trim=20 10 50 90, clip=true, scale = 0.26]{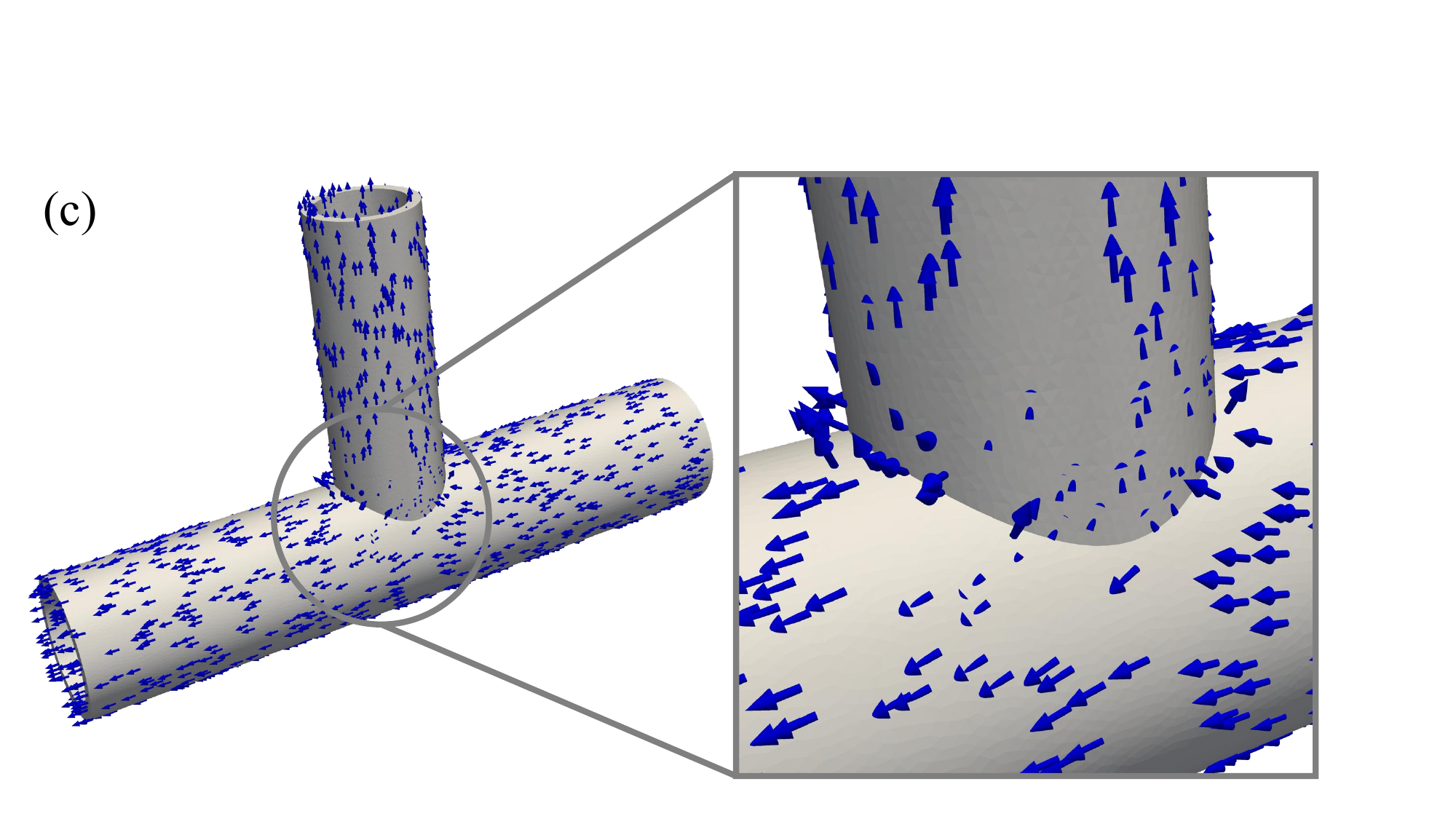} & \includegraphics[angle=0, trim=20 10 50 90, clip=true, scale = 0.26]{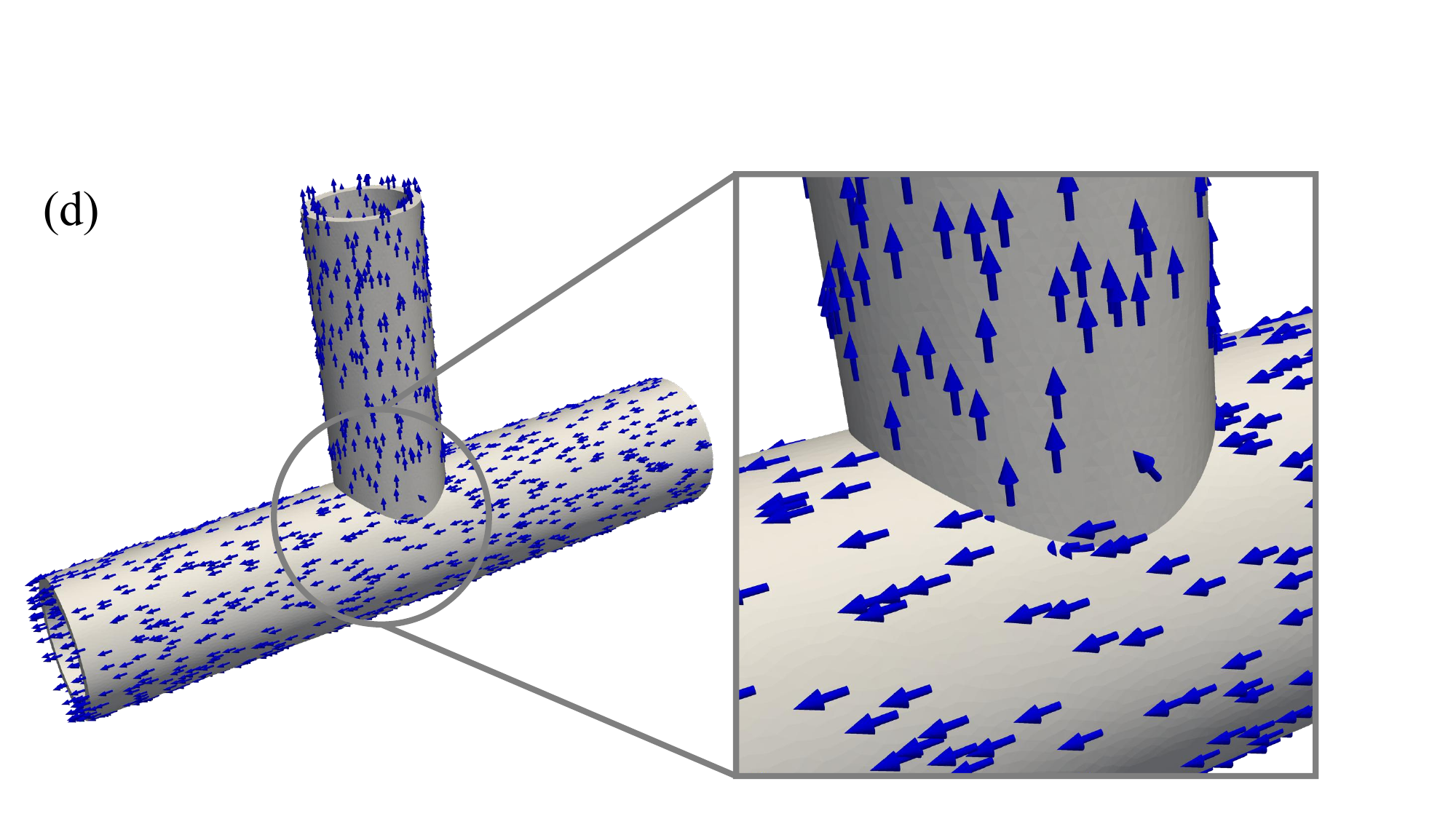} \\
\end{tabular}
\end{center}
\caption{(a) Circumferential direction given by the linear elastic analysis; (b) circumferential direction given by the morphology-based method; (c) axial direction given by the linear elastic analysis; (d) axial direction given by the morphology-based method}
\label{fig:local_basis_idealized}
\end{figure}

\subsection{Evaluation of methods for generating local basis vectors}
\label{sec:Evaluation_local_basis}
In this section, we perform a comparison between the approach for generating local basis vectors presented in Section \ref{sec:Local_basis} with the method described in \cite{Alastrue2010, Misiulis2019}. The latter employs static linear elastic analysis to generate local basis vectors. In specific, the geometry is loaded with internal pressure, and a Robin boundary condition is applied on the exterior surface to prevent bending and rigid motion. The principal directions of the maximum, mid, and minimum principal stresses define the local circumferential, axial, and radial directions of the vessel, respectively.

The comparison is conducted using a T-shaped geometry. The local circumferential and axial directions based on the elastic stresses are illustrated in Figure \ref{fig:local_basis_idealized} (a) and (c), respectively. In Figure \ref{fig:local_basis_idealized} (b) and (d), we present the local circumferential and axial vectors obtained using the proposed method. Both methods yield satisfactory local circumferential directions with minor differences. Nevertheless, for the local radial direction near the bifurcation region, the direction of the mid-principal stress noticeably differs from the actual axial direction. In contrast, our morphology-based approach robustly delivers higher-quality definition of the local axial direction.

\begin{figure}[htbp]
\begin{center}
\begin{tabular}{cc}
\includegraphics[angle=0, trim=190 100 180 60, clip=true, scale=0.4]{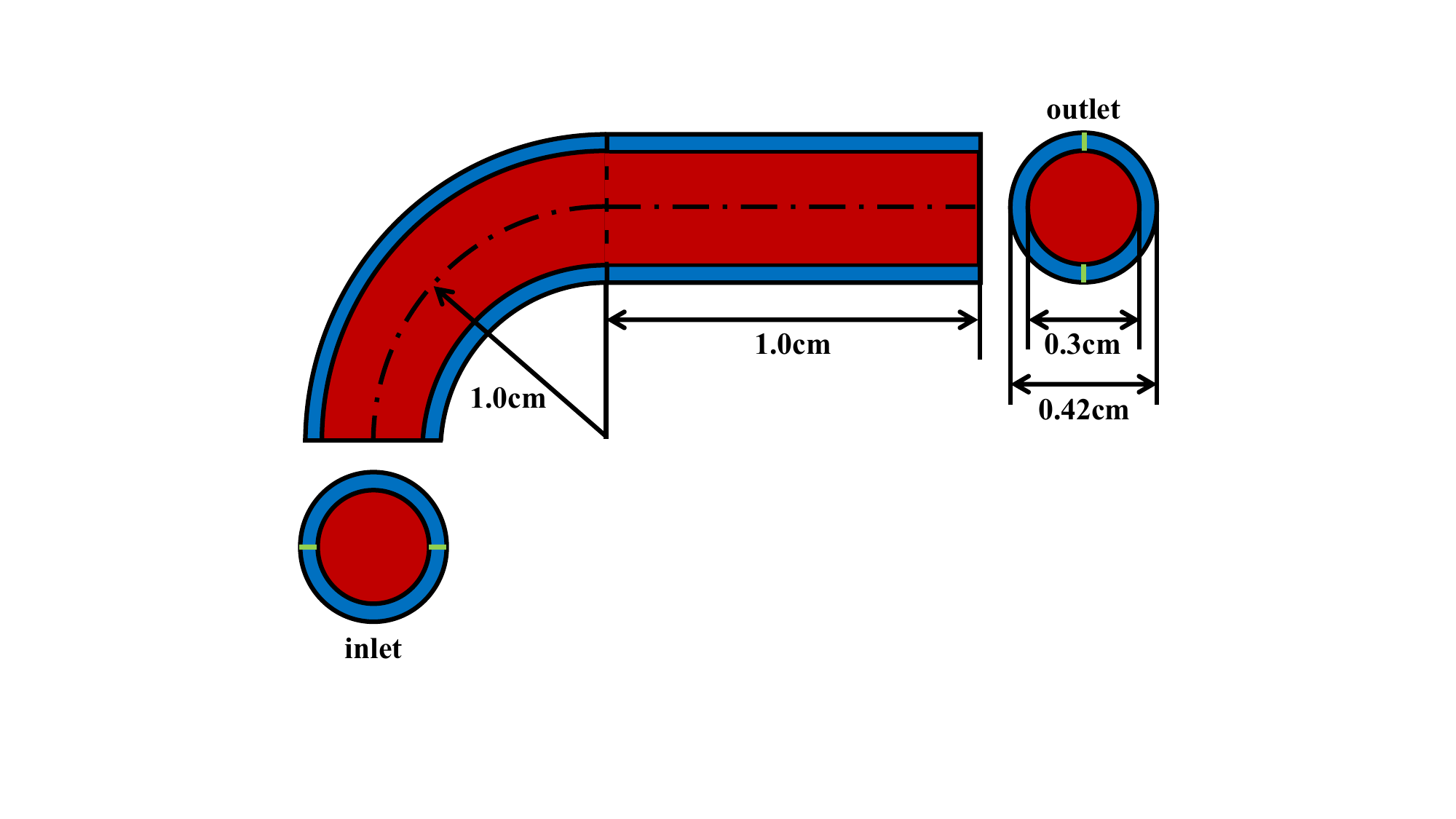} &
\includegraphics[angle=0, trim=150 10 150 120, clip=true, scale=0.18]{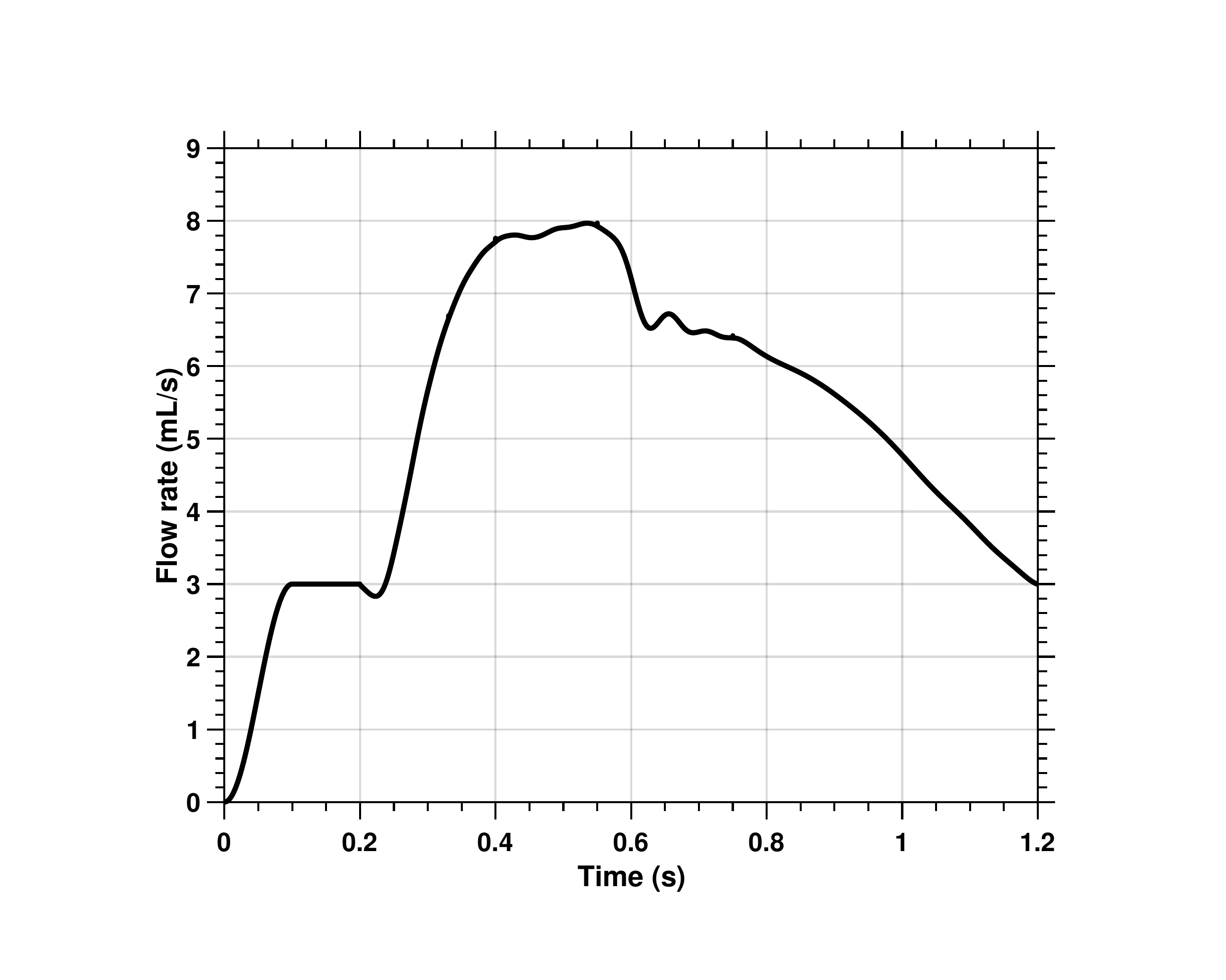}
\end{tabular}
\end{center}
\caption{Geometric setting (left) and the inflow profile over time (right) of the curved tube FSI benchmark}
\label{fig:CA-benchmark-geo-inflow}
\end{figure}

\begin{figure}[htbp]
	\begin{center}
		\includegraphics[width=1.0\linewidth, angle=0, trim=70 170 50 150, clip=true]{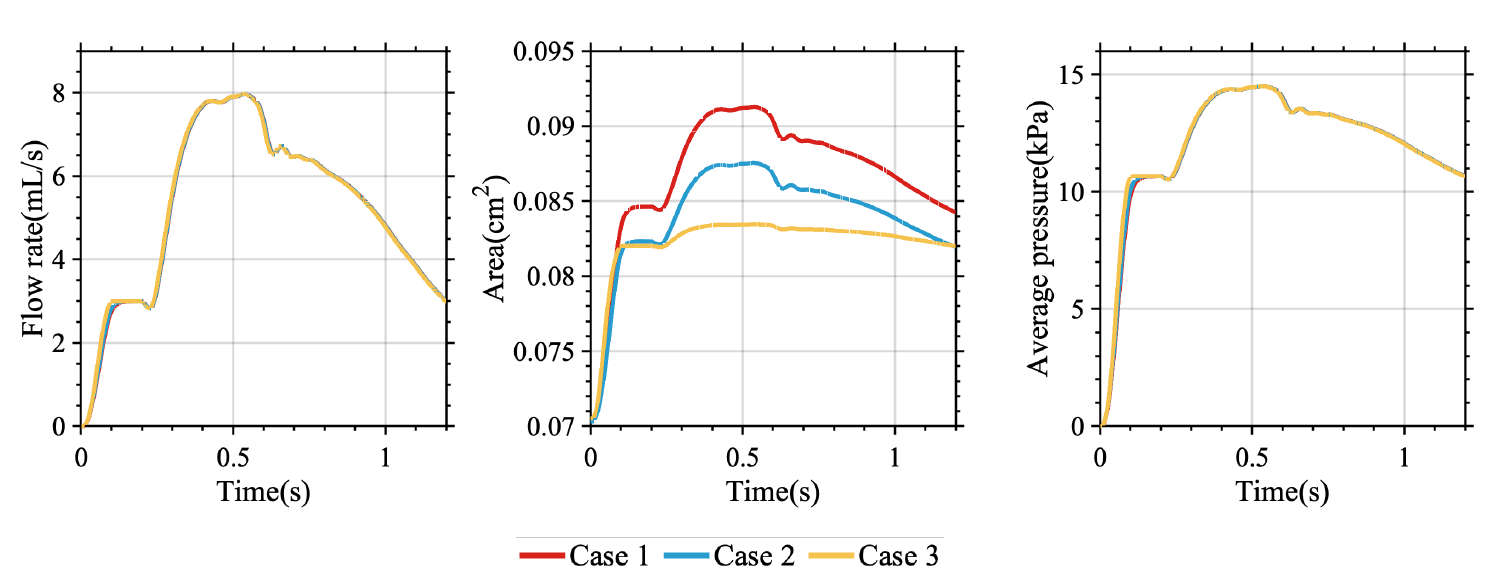}
	\end{center}
	\caption{The change of flow rate, area, and pressure of outlet.}
	\label{fig:3-case-PQA-CA-benchmark}
\end{figure}

\begin{figure}[htbp]
	\begin{center}
		\includegraphics[width=1.0\linewidth, angle=0, trim=00 160 00 160, clip=true]{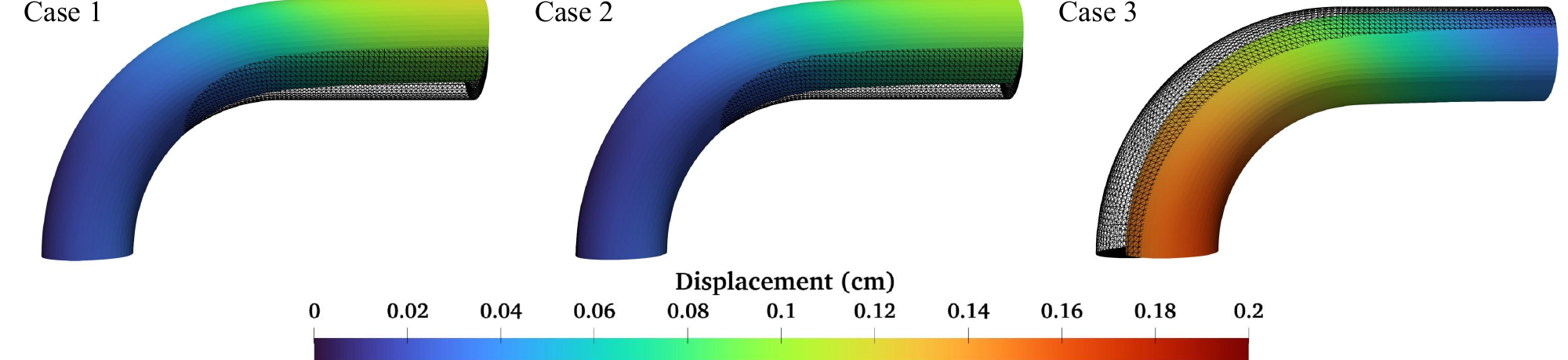}
	\end{center}
	\caption{The deformation of the arterial wall at $t = 0.536$ s.}
	\label{fig:NH-GOH-disp-5360}
\end{figure}

\begin{figure}[htbp]
	\begin{center}
		\includegraphics[width=1.0\linewidth, angle=0, trim=00 160 00 160, clip=true]{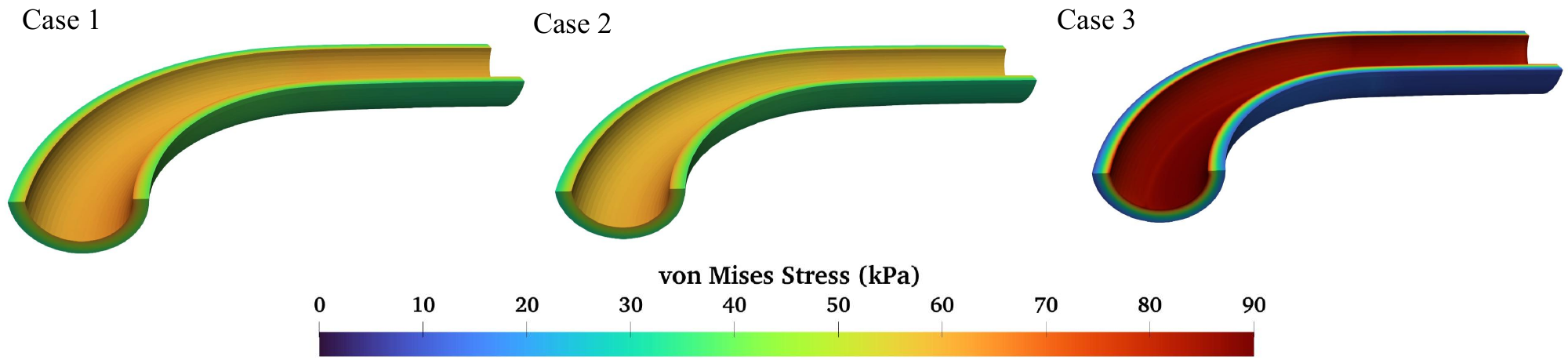}
	\end{center}
	\caption{The distribution of the von Mises stress in the arterial wall at $t = 0.536$ s.}
	\label{fig:NH-GOH-stress-5360}
\end{figure}

\subsection{Curved tube FSI benchmark}
\label{sec:Curved_tube_FSI}
We consider the benchmark proposed in \cite{Balzani2016,Balzani2023} to investigate the fiber reinforced material model in the setting of FSI. The geometric setting is illustrated in Figure \ref{fig:CA-benchmark-geo-inflow}, representing an idealized coronary artery. The axial displacement components of all nodes on the inlet and outlet annular surfaces are fixed, and the nodes marked by the green line are only allowed to move in the radial direction. On the inlet surface, the fluid velocity is prescribed with a parabolic profile with the flow rate over time shown in Figure \ref{fig:CA-benchmark-geo-inflow}. In this example, we do not have the prestress field initialized for the tissue. Instead, the curved elastic tube is inflated with a physiological pressure applied on the interior surface. This is referred to as the \textit{prestretch} procedure \cite{Balzani2016}. To have the tube inflated in a robust manner, the flow rate increases from zero to $3.0$ mL/s in $0.1$ s using a cosine-type ramp function of time and maintains at that value for another $0.1$ s. The chosen cosine-type ramp function assists in avoiding undesirable oscillations \cite{Balzani2016}. On the outlet surface of the fluid subdomain, a resistance boundary condition is applied with the resistance value being $35553.33$ dyn$\cdot$s/cm$^5$ and zero distal pressure, which gives a pressure of $80$ mmHg at the flow rate of $3.0$ mL/s. At time $t=0.2$ s, the flow and elastic tube can be viewed to reach a steady state, and the pulsatile flow profile is turned on. As is shown in Figure \ref{fig:CA-benchmark-geo-inflow}, the maximum flow rate reaches $7.97$ mL/s at time $t=0.536$ s, and the cardiac cycle is $1.0$ s. This pulsatile flow profile mimics that of a right coronary artery. At the outlet of the fluid subdomain, a resistance boundary condition is applied with the resistance and distal pressure values being $7720.86$ dyn$\cdot$s/cm$^5$ and $83472.39$ dyn/cm$^2$, which gives the outlet pressure of $80$ mmHg at $t=0.2$ and $108$ mmHg at $t=0.536$. The fluid density and dynamic viscosity are set to be 1.0 g/cm$^3$ and 0.04 dyn·s/cm$^2$, respectively. In the numerical study, we use a mesh with $5.18 \times 10^{5}$ linear tetrahedral elements, consisting of $3.36 \times 10^{5}$ solid elements and $ 1.82 \times 10^{5}$ fluid elements. Time integration is performed with a fixed time step size $\Delta t = 1.0\times10^{-4}$ s. We mention that a mesh independence study has been performed, and the reported mesh and time step sizes are the finest spatiotemporal resolution we used. 

\subsubsection{The curved tube fluid–structure benchmark problem}
\label{sec:CA-benchmark}
We first investigate the impact of the constitutive models of the elastic wall, and the results are compared with those reported in \cite{Balzani2023} for verification purposes. Three different material models are considered:
\begin{enumerate}
	\item[] Case 1: neo-Hookean model \eqref{eq:NH constitutive model} with $\mu = 127.52$ kPa and $K = 6333.3$ kPa.
	\item[] Case 2: neo-Hookean model \eqref{eq:NH constitutive model} with $\mu = 147.65$ kPa and $K = 7333.3$ kPa.
	\item[] Case 3: GOH model \eqref{eq:HGO-C constitutive model} with $\mu = 21.4$ kPa, $k_1 = 1018.8$ kPa, $k_2 = 20.0$, $\kappa = 0.0$, $\theta = 47^{\circ}$, and $K= 207.1$ kPa
\end{enumerate}
The material parameters for the GOH model are fitted based on a uniaxial tension test of the arterial media layer \cite{Brands2008,Holzapfel2006}, while the shear moduli for Cases 1 and 2 are chosen to approximate the stiffness of the experimental data at sufficiently large stretches \cite[Figure~1]{Balzani2023}. For the neo-Hookean model with even lower shear modulus, we do observe non-robust numerical behavior, consistent with the observation reported in \cite{Balzani2023}. 

Figure \ref{fig:3-case-PQA-CA-benchmark} illustrates the flow rate, area, and averaged pressure over time calculated on the outflow surface. The results match well with those of \cite{Balzani2016}. The flow rate and averaged pressure of these three models are very close. This is expected because the former is dictated by the global mass conservation, and the latter follows the imposed boundary condition. The area changes or the displacement on the outlet annular surface are larger for the neo-Hookean models, possibly due to the lack of fiber reinforcement.

Figures \ref{fig:NH-GOH-disp-5360} and \ref{fig:NH-GOH-stress-5360} depict the displacement and von Mises stress of the arterial wall at time $t = 0.536$ s, respectively. The black grid in Figure \ref{fig:NH-GOH-disp-5360} represents the initial position of the elastic tube. A significant difference between the isotropic and anisotropic models is clearly observable, while the differences between the two isotropic models are difficult to discern. The displacement magnitude of the neo-Hookean models is significantly smaller. In the meantime, the neo-Hookean tube bends upward near the outlet, while the GOH bends medially near the inlet. The anisotropic model exhibits a localized stress distribution, suggesting the fibers are sufficiently stretched near the luminal interface and carry a significant amount of loads. 

\begin{figure}[htbp]
	\begin{center}
		\includegraphics[width=0.9\linewidth, angle=0, trim=0 10 0 0, clip=true]{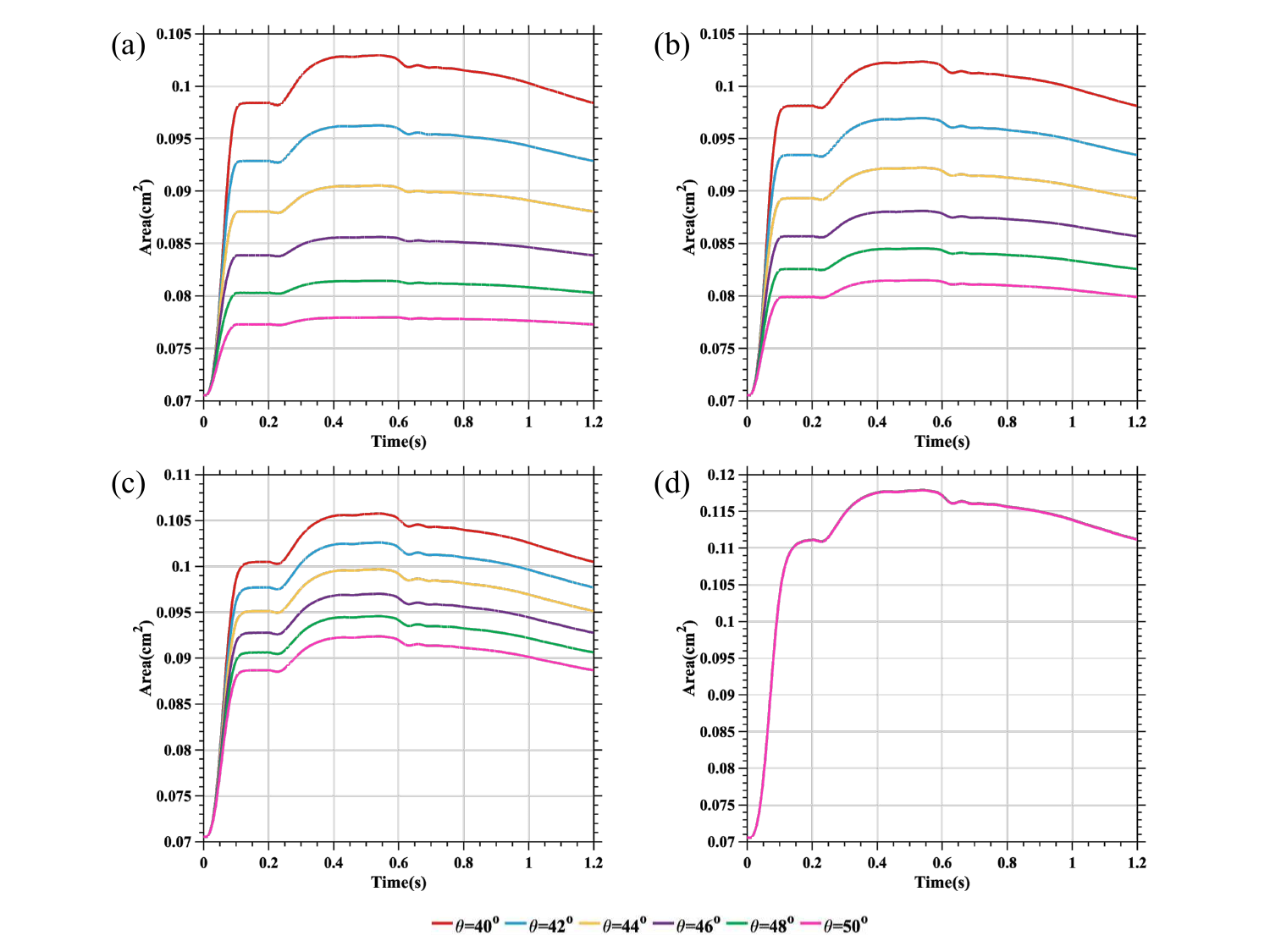}
	\end{center}
	\centering
	\caption{Area change of outlet: (a) $\kappa = 0.0$, (b) $\kappa = 0.111$, (c) $\kappa = 0.222$, (d) $\kappa = 0.333$.}
	\label{fig:8case-GOH-area-chage}
\end{figure}

\begin{figure}[htbp]
	\begin{center}
		\includegraphics[width=1.0\linewidth, angle=0, trim=0 130 40 80, clip=true]{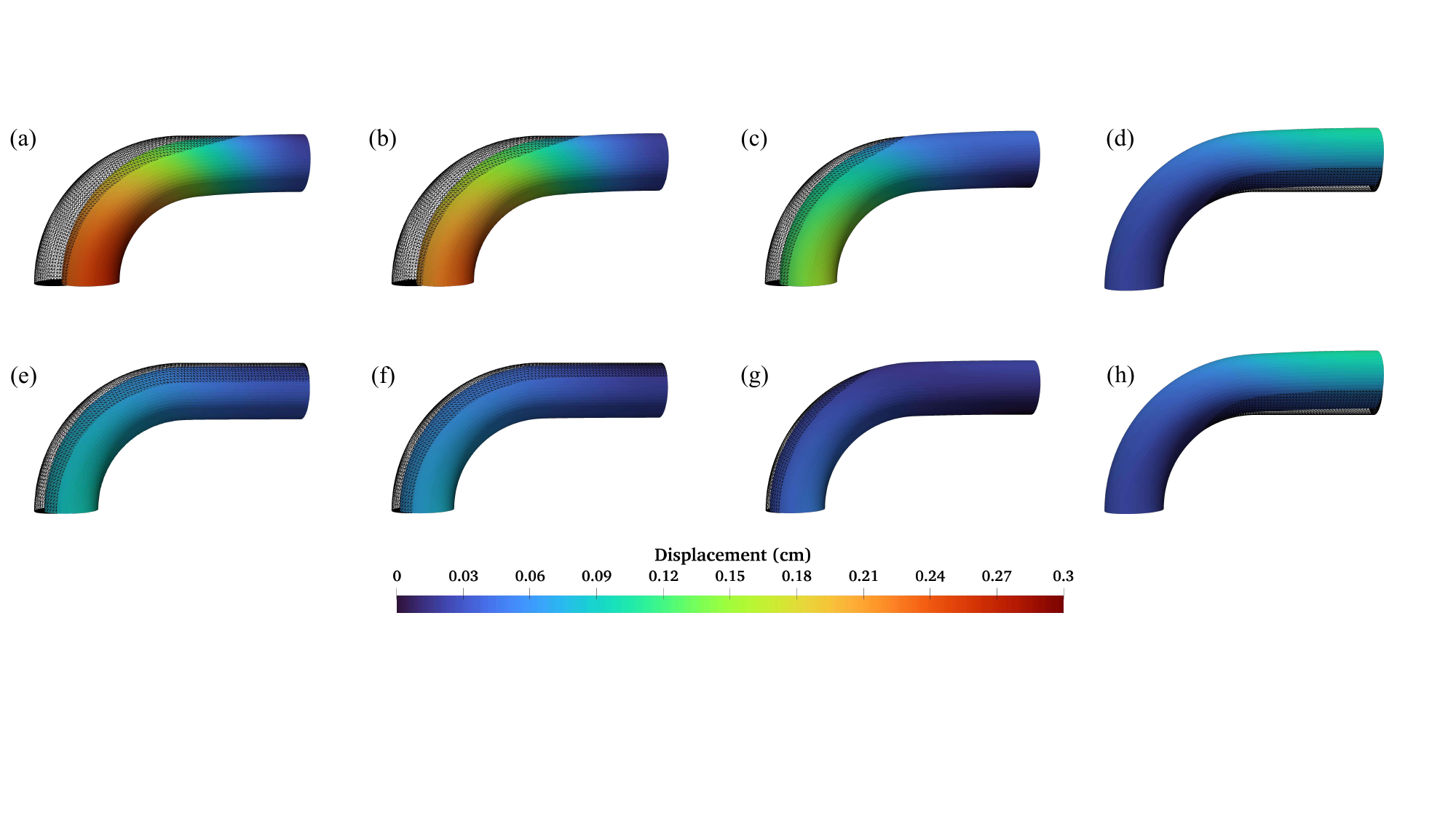}
	\end{center}
	\centering
	\caption{Displacement of the arterial wall at $t = 0.536$ s: (a) $\kappa = 0.0, \theta = 40^{\circ}$, (b) $\kappa = 0.111, \theta = 40^{\circ}$, (c) $\kappa = 0.222, \theta = 40^{\circ}$, (d) $\kappa = 0.333, \theta = 40^{\circ}$, (e) $\kappa = 0.0, \theta = 50^{\circ}$, (f) $\kappa = 0.111, \theta = 50^{\circ}$, (g) $\kappa = 0.222, \theta = 50^{\circ}$, (h) $\kappa = 0.333, \theta = 50^{\circ}$.}
	\label{fig:8case-GOH-disp-5360}
\end{figure}

\begin{figure}[htbp]
	\begin{center}
		\includegraphics[width=1.0\linewidth, angle=0, trim=0 90 10 130, clip=true]{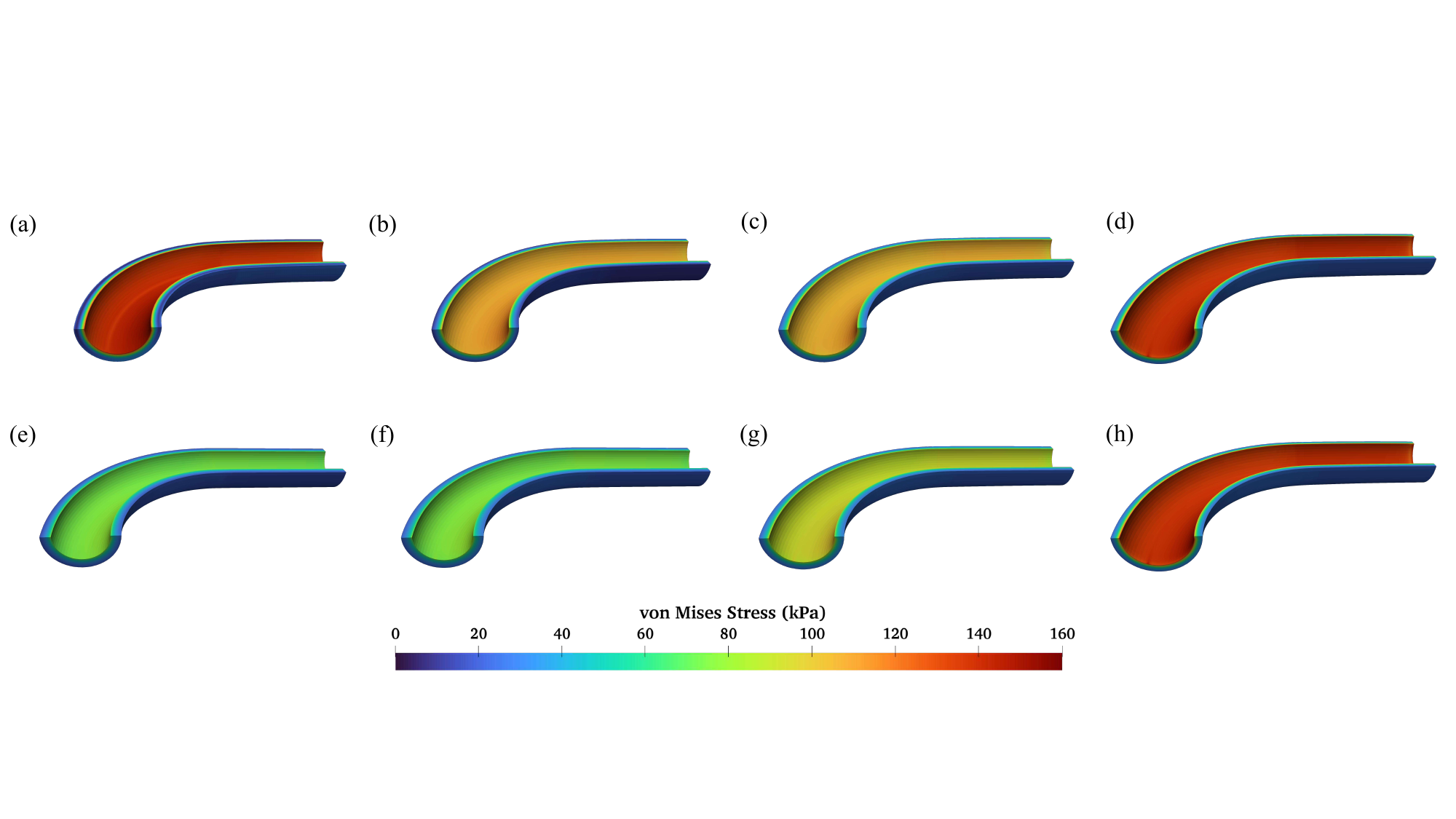}
	\end{center}
	\centering
	\caption{Von Mises stress distribution of the arterial wall at $t = 0.536$ s: (a) $\kappa = 0.0, \theta = 40^{\circ}$, (b) $\kappa = 0.111, \theta = 40^{\circ}$, (c) $\kappa = 0.222, \theta = 40^{\circ}$, (d) $\kappa = 0.333, \theta = 40^{\circ}$, (e) $\kappa = 0.0, \theta = 50^{\circ}$, (f) $\kappa = 0.111, \theta = 50^{\circ}$, (g) $\kappa = 0.222, \theta = 50^{\circ}$, (h) $\kappa = 0.333, \theta = 50^{\circ}$.}
	\label{fig:8case-GOH-stress-5360}
\end{figure}

\begin{figure}[htbp]
	\begin{center}
		\includegraphics[width=1.0\linewidth, angle=0, trim=0 90 10 130, clip=true]{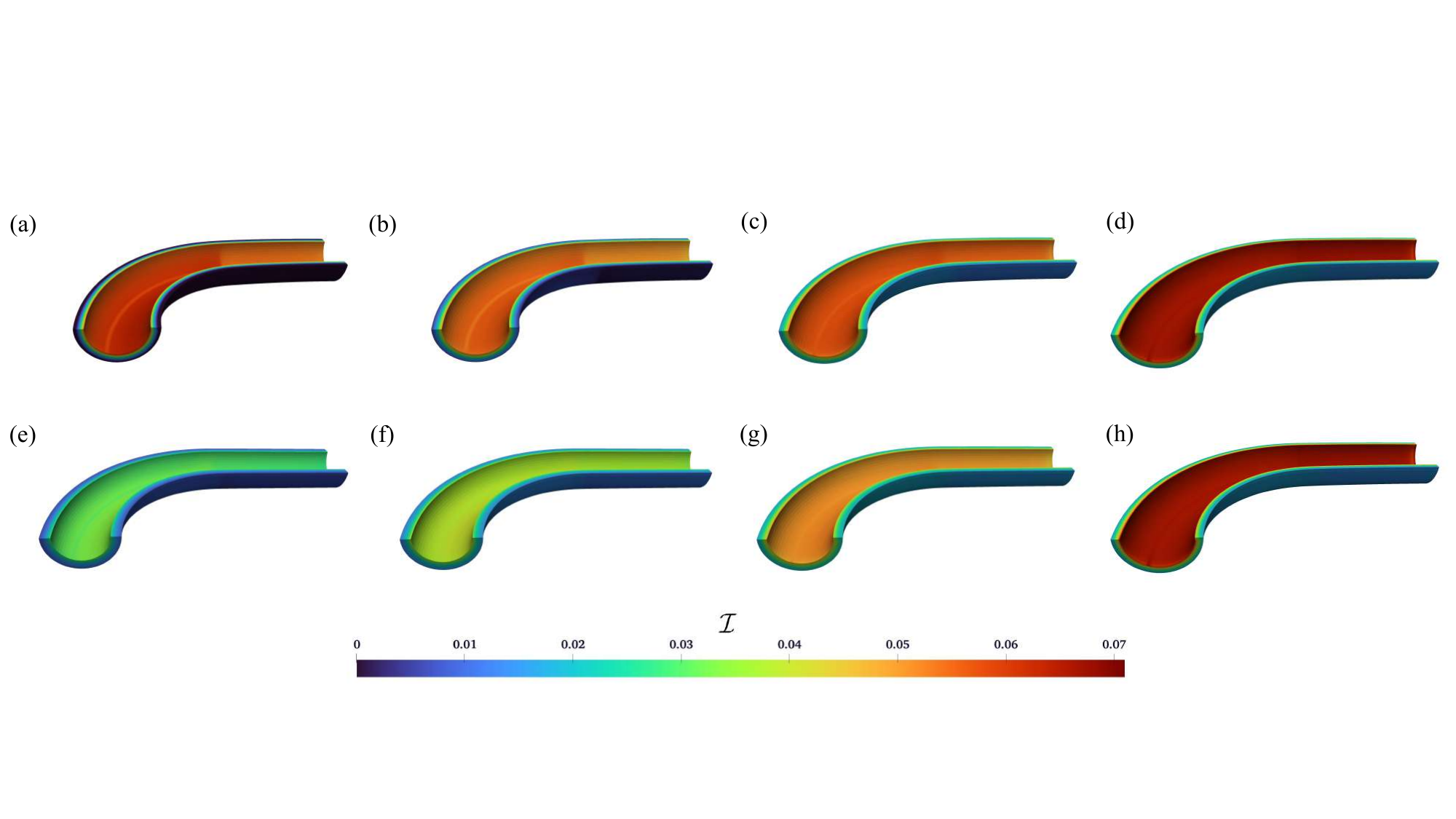}
	\end{center}
	\centering
	\caption{The invariant $\mathcal{I}$ distribution of the arterial wall at $t = 0.536$ s: (a) $\kappa = 0.0, \theta = 40^{\circ}$, (b) $\kappa = 0.111, \theta = 40^{\circ}$, (c) $\kappa = 0.222, \theta = 40^{\circ}$, (d) $\kappa = 0.333, \theta = 40^{\circ}$, (e) $\kappa = 0.0, \theta = 50^{\circ}$, (f) $\kappa = 0.111, \theta = 50^{\circ}$, (g) $\kappa = 0.222, \theta = 50^{\circ}$, (h) $\kappa = 0.333, \theta = 50^{\circ}$.}
	\label{fig:8case-GOH-EXP-5360}
\end{figure}

\subsubsection{Parameter test of GOH material model}
In this section, the GOH model is further investigated by inspecting the impact of the fiber orientation and dispersion. The problem settings remain the same as those in Section \ref{sec:CA-benchmark}. The values of $\theta$ are taken as $40^{\circ}$, $42^{\circ}$, $44^{\circ}$, $46^{\circ}$, $48^{\circ}$, and $50^{\circ}$, while the values of $\kappa$ are set to $0.0$, $0.111$, $0.222$, and $0.333$.

The changes in the area over time at the outlet are shown in Figure \ref{fig:8case-GOH-area-chage}. It can be observed that under a fixed value of $\kappa$, the area change becomes more pronounced when the angle $\theta$ is smaller, meaning the fibers are more aligned with the axial direction. In the considered FSI problem, the deformation behavior of the vessel wall predominantly involves radial expansion caused by the flow pressure. Therefore, when the fiber orientation is closer to the circumferential direction, it more effectively activates fiber reinforcement. Moreover, the results show that the differences between the curves of different angle values diminish as $\kappa$ increases. This is because larger $\kappa$ values indicate a more dispersed fiber orientation, rendering the impact of $\theta$ less significant. In particular, when $\kappa=0.333$, the material becomes isotropic and the angle $\theta$ no longer affects the material behavior.

Figure \ref{fig:8case-GOH-disp-5360} shows the displacement distribution of the wall at time $t = 0.536$ s, with the black grid representing the initial position of the wall. The results of $\theta=40^{\circ}$ and $\theta=50^{\circ}$ are depicted. When comparing the results of different angle values with the same $\kappa$ value, the displacements of $\theta=40^{\circ}$ are consistently larger than those of $\theta=50^{\circ}$. This is because the fibers oriented with $\theta = 50^{\circ}$ provide more reinforcement of the wall. When comparing the results of different $\kappa$ values with the same angle, the wall displacement decreases with the increase of the fiber dispersion. Interestingly, the deformation of the wall becomes more similar to that of the neo-Hookean model when $\kappa$ gets closer to $0.333$. It can be observed that wall bends upward near the outlet and the displacement becomes negligible near the inlet when $\kappa=0.333$. Conversely, when $\kappa = 0.0$, the wall bends medially near the inlet with almost no displacement at the outlet. This is consistent with the prior observation made in \cite{Balzani2023} and Figure \ref{fig:NH-GOH-disp-5360}, indicating that the deformation state of Case 3 in Figure \ref{fig:NH-GOH-disp-5360} is in fact caused by the anisotropy rather than the exponential energy form. The differences in the deformation states suggest that considering anisotropy in the tissue model may lead to substantially different behavior under physiological loading.

Figure \ref{fig:8case-GOH-stress-5360} further illustrates the von Mises stress distribution at time $t = 0.536$ s. When the angle $\theta = 50^{\circ}$, the von Mises stress increases with the increase of $\kappa$ value. However, when $\theta = 40^{\circ}$, the relation between the von Mises stress and the dispersion parameter is less straightforward. The stress initially decreases and then increases as $\kappa$ increases from $0.0$ to $0.333$. To gain an understanding of this phenomenon, we monitor the values of $\mathcal{I}:= \kappa{I}_{1}+(1-3\kappa){I}_4-1$, as shown in Figure \ref{fig:8case-GOH-EXP-5360}. The term $\mathcal I$ appears in the exponential function of the strain energy \eqref{eq:HGO-C constitutive model}, and its variation pattern with respect to $\kappa$ mirrors that of the von Mises stress for both angles. The value of $\mathcal I$ is influenced by the combined effects of the dispersion parameter and the deformation state, and this explains the von Mises stresses illustrated in Figure \ref{fig:8case-GOH-stress-5360}.

\subsection{Patient-specific abdominal aorta model}
\label{sec:AA model FSI}
\begin{figure}[htbp]
	\begin{center}
		\includegraphics[angle=0, trim=30 80 60 65, clip=true, scale = 0.5]{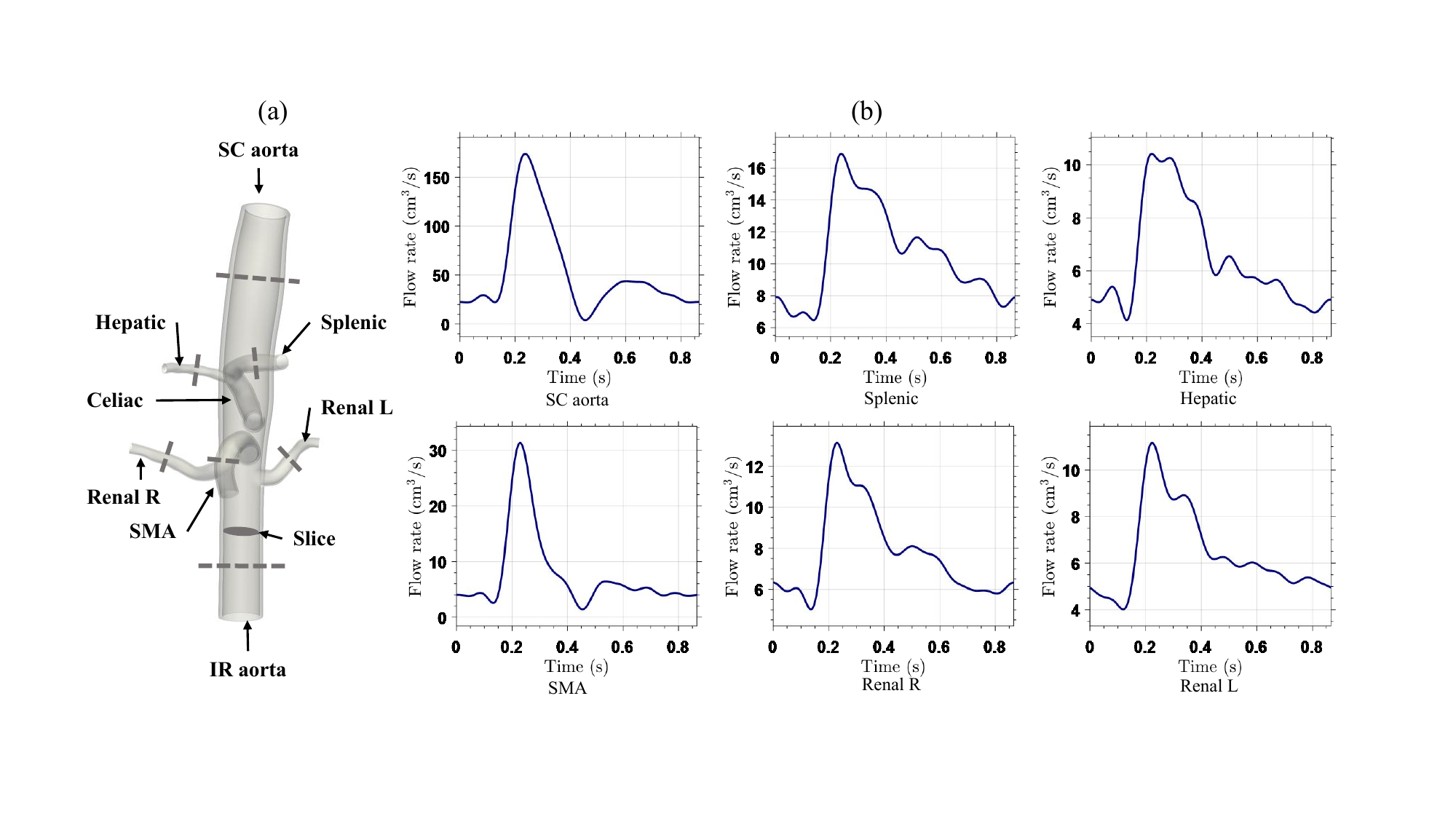}
	\end{center}
	\caption{ (a) The geometry of AA model. (b) The flow waveform of the supra-celiac aorta inlet and the outlets of visceral arteries.}
	\label{fig:AA_diagram1}
\end{figure}

In this example, we consider an image-based abdominal aorta (AA) model, which includes multiple visceral arteries (hepatic, splenic, renal, and superior mesenteric arteries (SMA)). The geometric model is constructed based on the MRI data from a healthy adult male \cite{Chen2022,Shi2021} (Figure \ref{fig:AA_diagram1} (a)). The non-contrast-enhanced magnetic resonance angiogram images are obtained using the inflow inversion recovery sequence on a 3.0 T magnetic resonance scanner (Discovery MR750, GE Medical Systems) equipped with a 32-channel coil. More details of this scanner can be found in \cite{Shi2021}.

The images at the cross-sections (illustrated by the dashed lines in Figure \ref{fig:AA_diagram1} (a)) of the visceral arteries, supra-celiac (SC) aorta, and infrarenal (IR) aorta are obtained using a pulse and respiratory gated 2D phase contrast sequence. The obtained 2D phase-contrast magnetic resonance imaging (PC-MRI) data are processed using Segment \cite{segment-website} to extract the physiological flow rates at these cross-sections for 30 time frames within a cardiac cycle. Polynomial interpolation and phase correction are applied to these flow rates to obtain the flow waveforms at the inlet and outlets through a cardiac cycle. Due to limitations in spatial resolution, the 2D PC-MRI technique is insufficient to accurately identify small arteries, such as the left gastric arteries and lumbar arteries, their flow rates are excluded. To ensure mass conservation, the sum of the flow waveforms of the IR aorta and all visceral arteries is set as the flow waveform for the inlet of the SC aorta. These obtained flow waveforms are then combined with a parabolic flow profile to prescribe the velocity at the inlet and the outlets of the five visceral arteries (Figure \ref{fig:AA_diagram1} (b)). 

\begin{figure}[htbp]
	\begin{center}
		\includegraphics[angle=0, trim=20 10 50 70, clip=true, scale = 0.45]{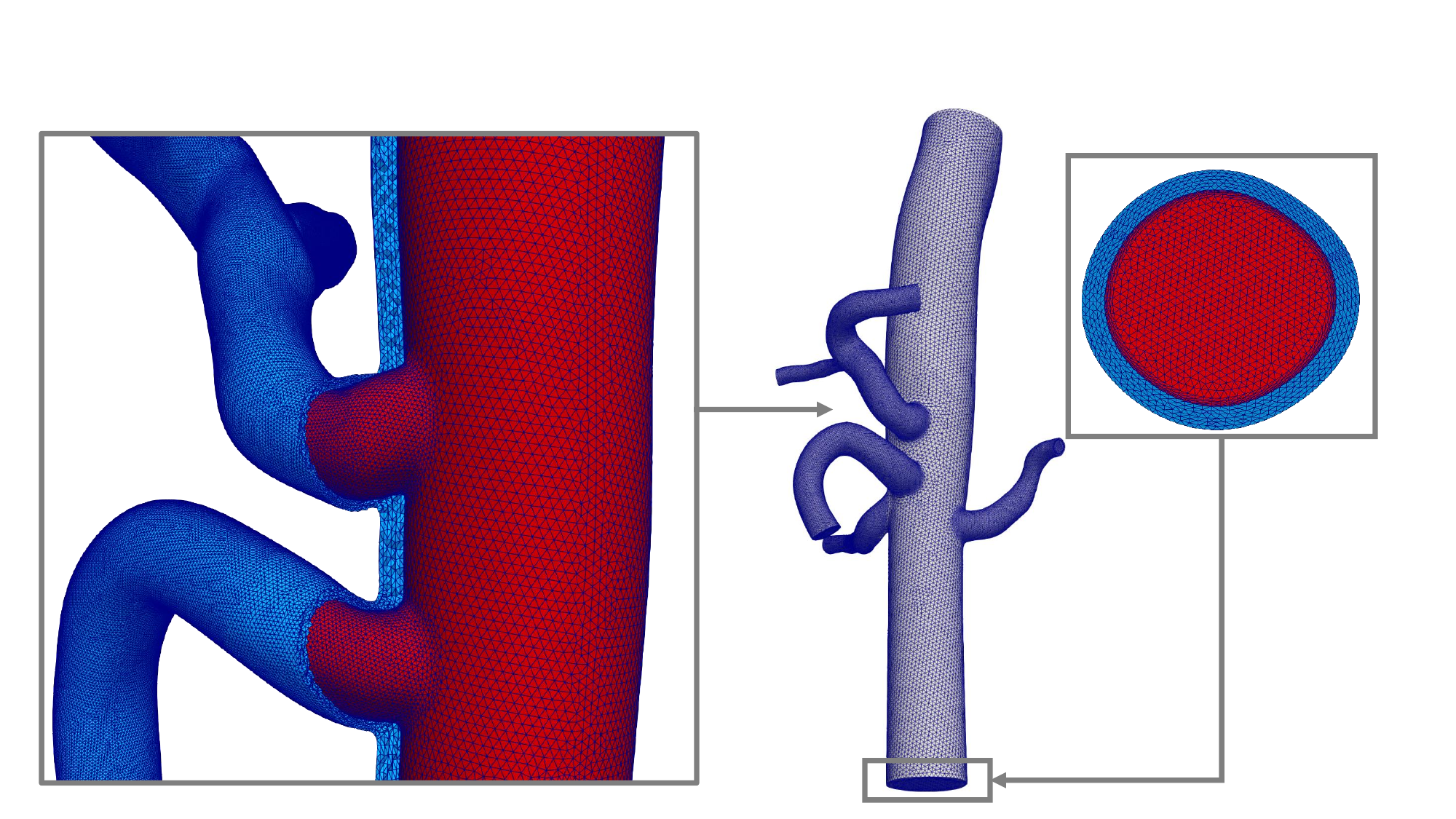}
	\end{center}
	\caption{The mesh of AA model.}
	\label{fig:mesh_AA_diagram1}
\end{figure}

\begin{figure}[htbp]
	\begin{center}
		\includegraphics[angle=0, trim=80 5 80 5, clip=true, scale = 0.5]{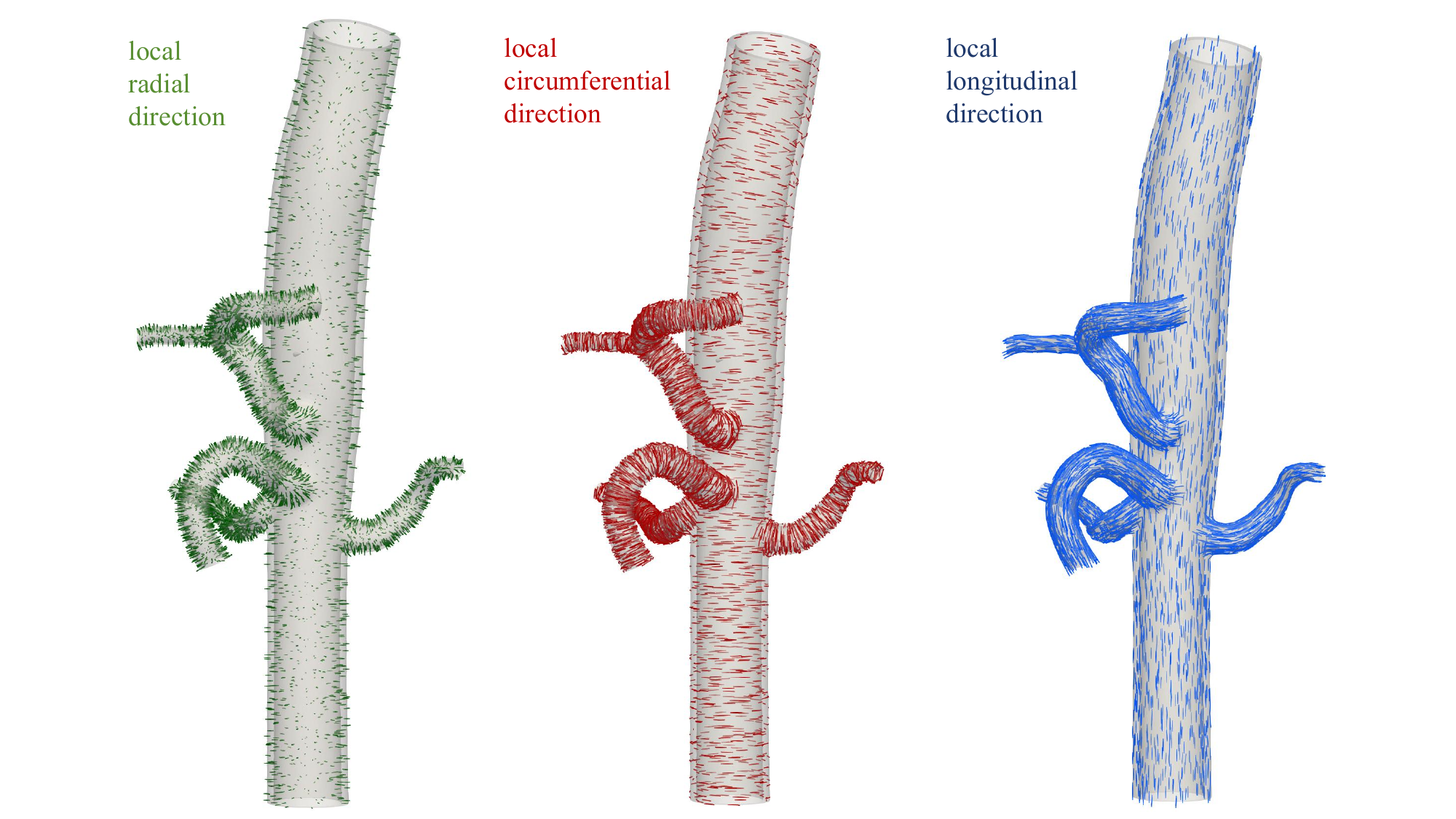}
	\end{center}
	\caption{The generated local basis vectors of AA model.}
	\label{fig:local_basis_AA_diagram1}
\end{figure}

Regarding the outlet of the IR aorta, a three-element Windkessel model is coupled with the 3D FSI model as the outflow boundary condition. Utilizing the measured flow waveform at the IR arterial outlet, the RCR parameters are tuned to ensure that the peak and trough of the output pressure match the systolic pressure ($P_\mathrm{sys} = 103$ mmHg) and diastolic pressure ($P_\mathrm{dia} = 72$ mmHg) of the subject. The specific settings are as follows. The compliance of the Windkessel model is set to 0.000452 cm$^3$/dyn; the proximal and distal resistances are set to 330.69 dyn$\cdot$s/cm$^3$ and 7708.28 dyn$\cdot$s/cm$^3$, respectively. In this study, the exterior wall of the artery is kept stress-free, meaning the supporting tissue is not considered. Meanwhile, on the inlet and outlet solid surfaces, the ring-shaped walls are fixed.

We employ the isotropic neo-Hookean model \eqref{eq:NH constitutive model} and the anisotropic GOH model \eqref{eq:HGO-C constitutive model} for the arterial wall to perform a comparative study. The solid material properties utilized are adopted from \cite{Balzani2023} and are summarized in Table \ref{table:aa_material_properties}. Moreover, the effect of fiber dispersion is investigated by considering two different dispersion parameters (i.e., $\kappa=0.0$ and $0.226$ \cite{Holzapfel2004a}). The blood density is set to 1.06 g/cm$^3$, and the dynamic viscosity is taken to be 0.04 dyn$\cdot$s/cm$^2$. The geometric model is discretized into $8.52 \times 10^6$ linear tetrahedral elements, including $4.64\times 10^6$ solid and $3.89 \times 10^6$ fluid elements. Three layers of BL elements with a growth ratio of 0.75 are constructed within the lumen. In the meantime, we strive to have at least six layers of elements across the vessel wall thickness (Figure \ref{fig:mesh_AA_diagram1}). We also investigate the problem with a relatively coarse mesh, which delivers essentially identical results. To construct the orientation of fibers, the local basis vectors are generated on the vessel wall using the procedures developed in Section \ref{sec:Local_basis} (Figure \ref{fig:local_basis_AA_diagram1}). 

\begin{table}[htbp]
	\small
	\begin{center}
		\tabcolsep=0.13cm
		\renewcommand{\arraystretch}{1.2}
		\begin{tabular}{c c c c c c c c c c c}
			\hline
			& Case &Model &$\rho_0$ (g$\cdot$cm$^{-3}$) &$K$ (kPa) & $\mu$ (kPa) & $\theta$ ($^{\circ}$) &$\gamma$ ($^{\circ}$) &$k_1$ (kPa) & $k_2$ & $\kappa$\\
			\hline
			&1 & GOH & $1.0$ &$676.20$ & $41.4$ & $47$ & $0$ & $ 2018.80$ & $ 20$ & $0.0$ \\
			&2 & GOH & $1.0$ &$676.20$ & $41.4$ & $47$ & $0$ & $ 2018.80$ & $ 20$ & $0.226$ \\
			&3 & neo-Hookean & $1.0$ & $4165.65$ & $255.04$ & - & - & - & - & - \\
			&4 & neo-Hookean & $1.0$ & $4823.23$ & $295.30$ & - & - & - & - & - \\
			\hline
		\end{tabular}
	\end{center}
	\caption{Material parameters used in the AA model. }
	\label{table:aa_material_properties}
\end{table}

\begin{figure}[htbp]
	\begin{center}
		\includegraphics[angle=0, trim=0 10 0 0, clip=true, scale = 0.45]{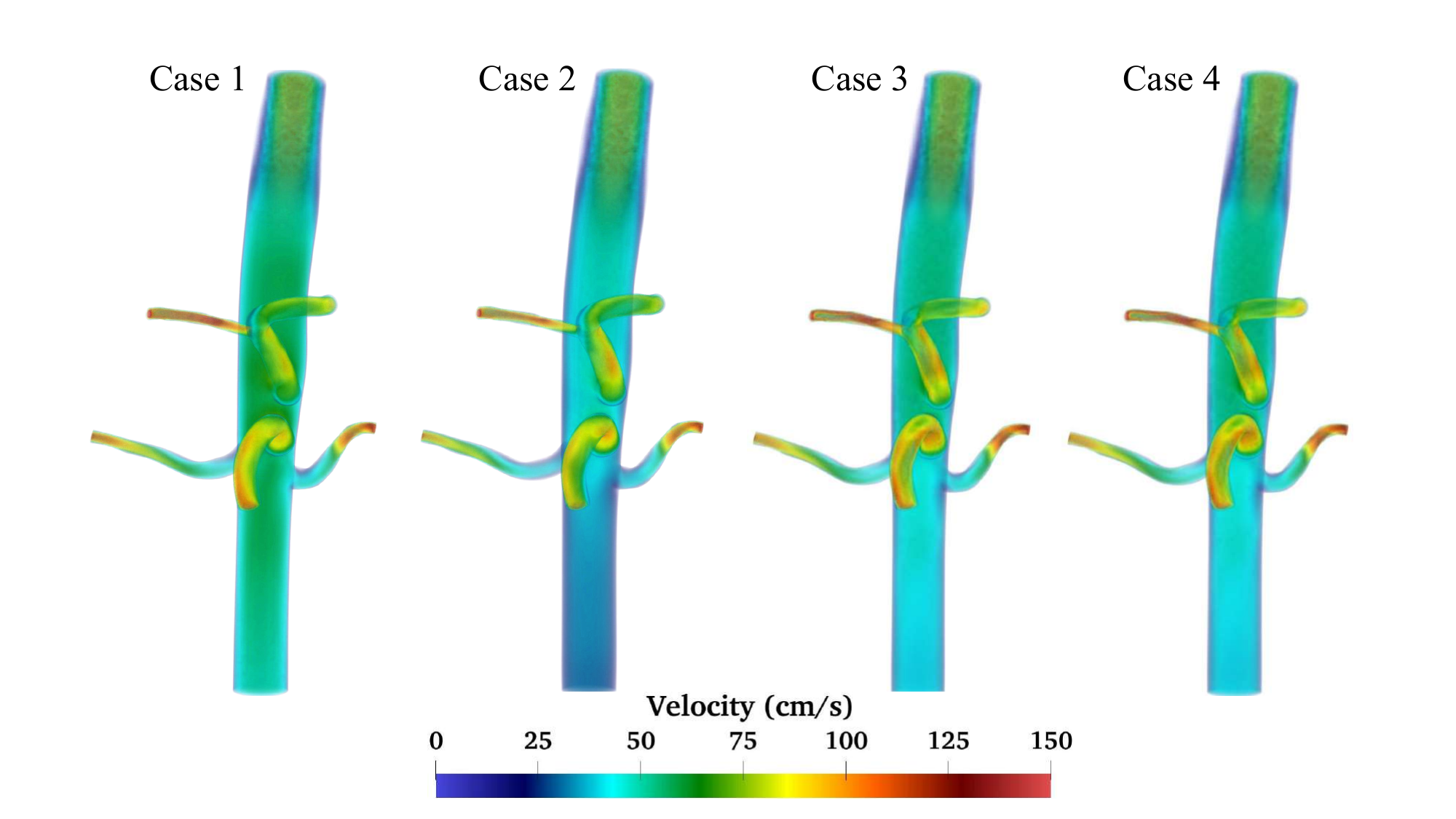}
	\end{center}
	\caption{The velocity computed at peak systole.}
	\label{fig:aa_velo_0.25s}
\end{figure}

\begin{figure}[htbp]
	\begin{center}
		\includegraphics[angle=0, trim=0 10 0 0, clip=true, scale = 0.45]{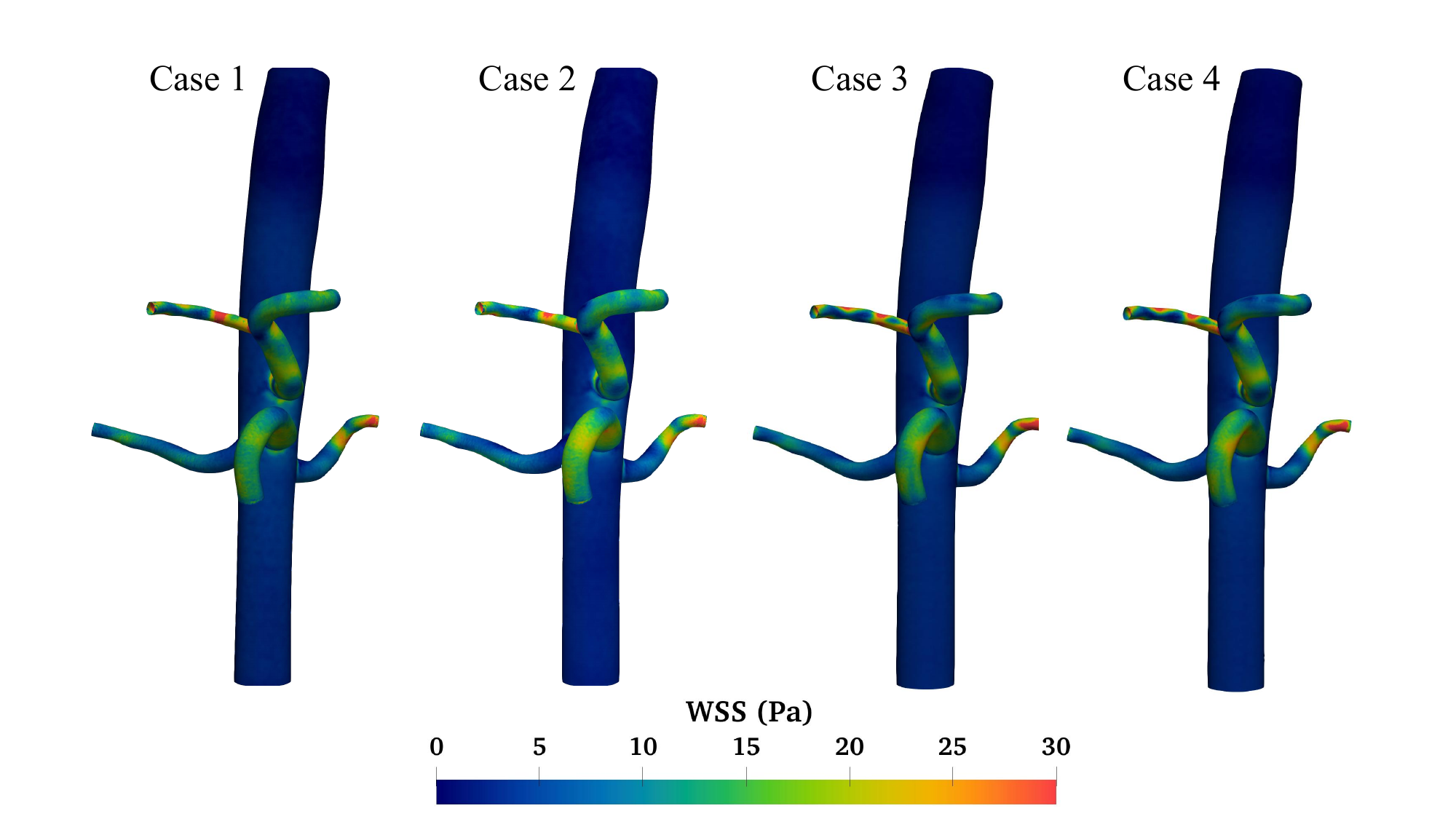}
	\end{center}
	\caption{The WSS computed at peak systole.}
	\label{fig:aa_wss_0.25s}
\end{figure}

FSI simulations are performed using the initialization procedure outlined in Section \ref{subsec:prestressing}. Two cardiac cycles are simulated with a fixed time step size of $\Delta t = 1.0\times10^{-5}$ s, and the results from the last cycle are presented here. Figures \ref{fig:aa_velo_0.25s} and \ref{fig:aa_wss_0.25s} depict the flow fields and WSS obtained by the four different material models at peak systole. It is evident that the flow and WSS distributions derived from these models are quite close, with only minor discrepancies in the flow field near the outlet of the IR aorta. 

\begin{figure}[htbp]
	\begin{center}
		\includegraphics[angle=0, trim=0 5 0 0, clip=true, scale = 0.45]{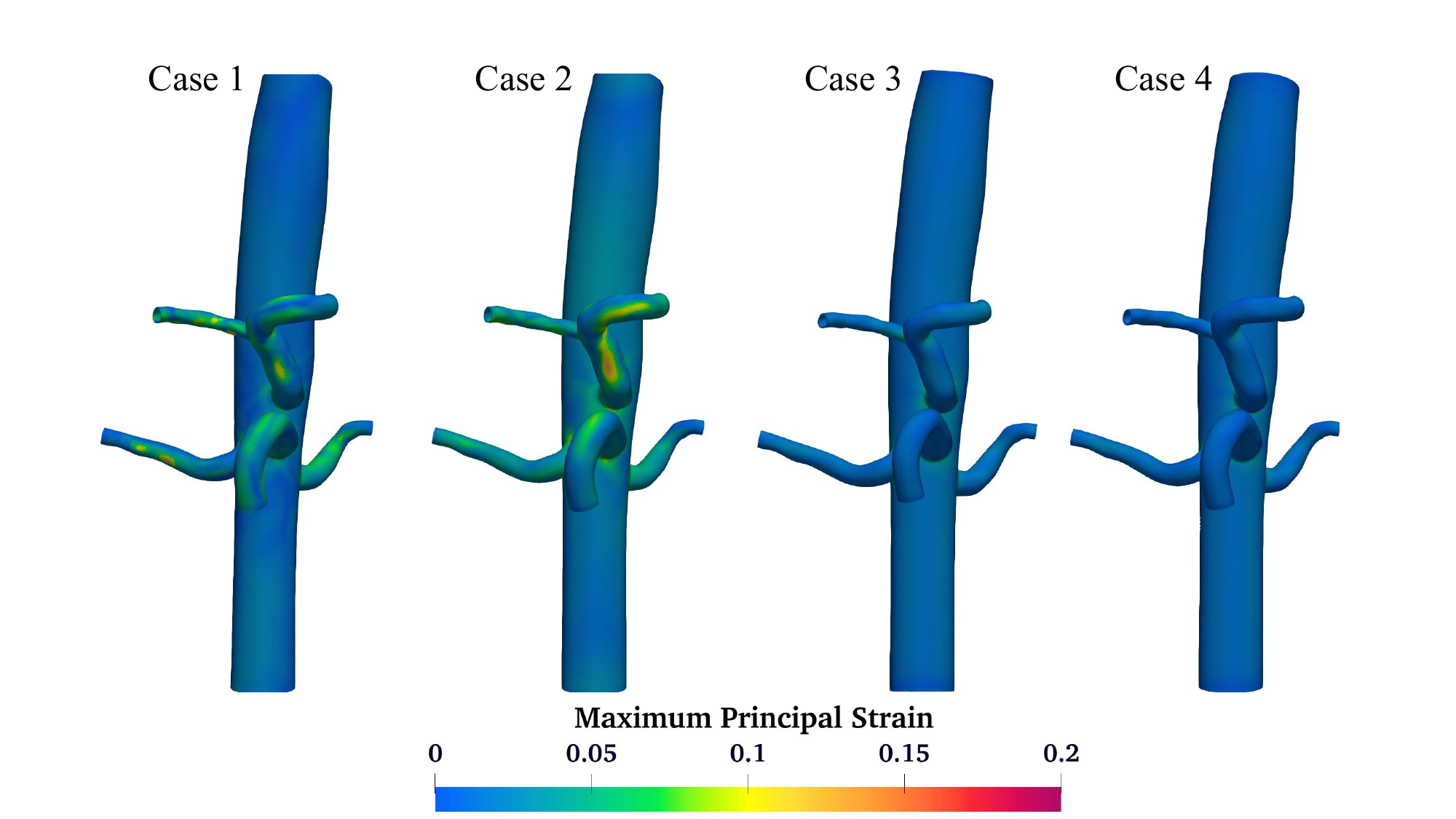}
	\end{center}
	\caption{The maximum principal strain of the arterial wall computed at peak systole.}
	\label{fig:aa_strain_0.25s}
\end{figure}

\begin{figure}[htbp]
	\begin{center}
		\includegraphics[angle=0, trim=0 10 0 0, clip=true, scale = 0.45]{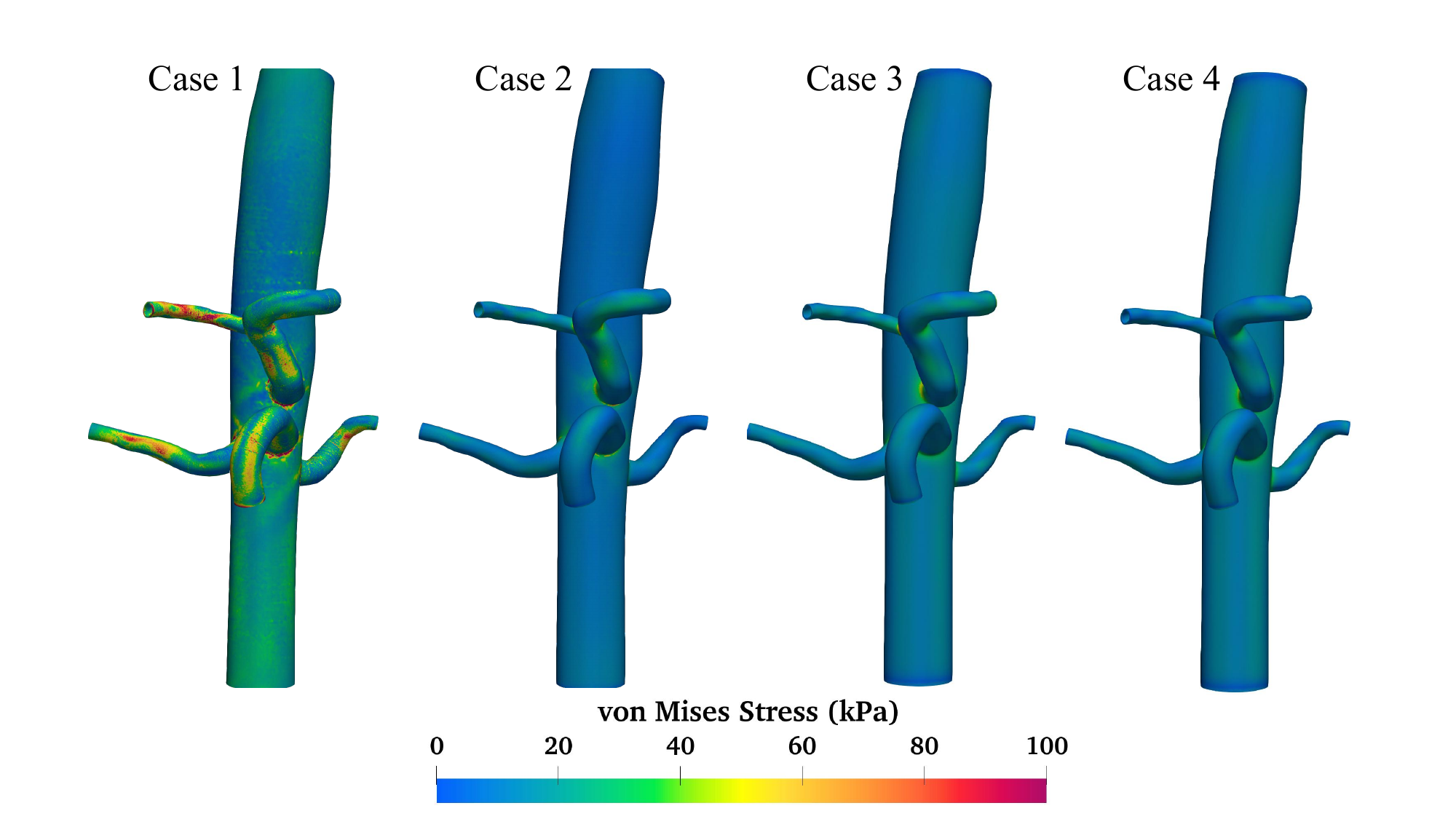}
	\end{center}
	\caption{The von Mises stress of the arterial wall computed at peak systole.}
	\label{fig:aa_stress_0.25s}
\end{figure}

Figures \ref{fig:aa_strain_0.25s} and \ref{fig:aa_stress_0.25s} depict the maximum principal strain and von Mises stress at peak systole for the four material models, respectively. For the isotropic models, the principal strain is lower and distributed smoothly within the wall; for the anisotropic models, the principal strain is higher in the visceral arteries. The stress pattern exhibits similarity between the two neo-Hookean models and the GOH model with $\kappa = 0.226$. Comparing the stress results for the GOH model with $\kappa = 0.0$ and $\kappa = 0.226$, the former exhibits a more localized stress distribution, with higher values observed at various branches of visceral arteries. Moreover, the stress of Case 1 is significantly higher than that in the rest cases. This discrepancy is due to the strong material anisotropy for Case 1.

\begin{figure}[htbp]
		\begin{center}
		\includegraphics[angle=0, trim=150 105 100 40, clip=true, scale = 0.3]{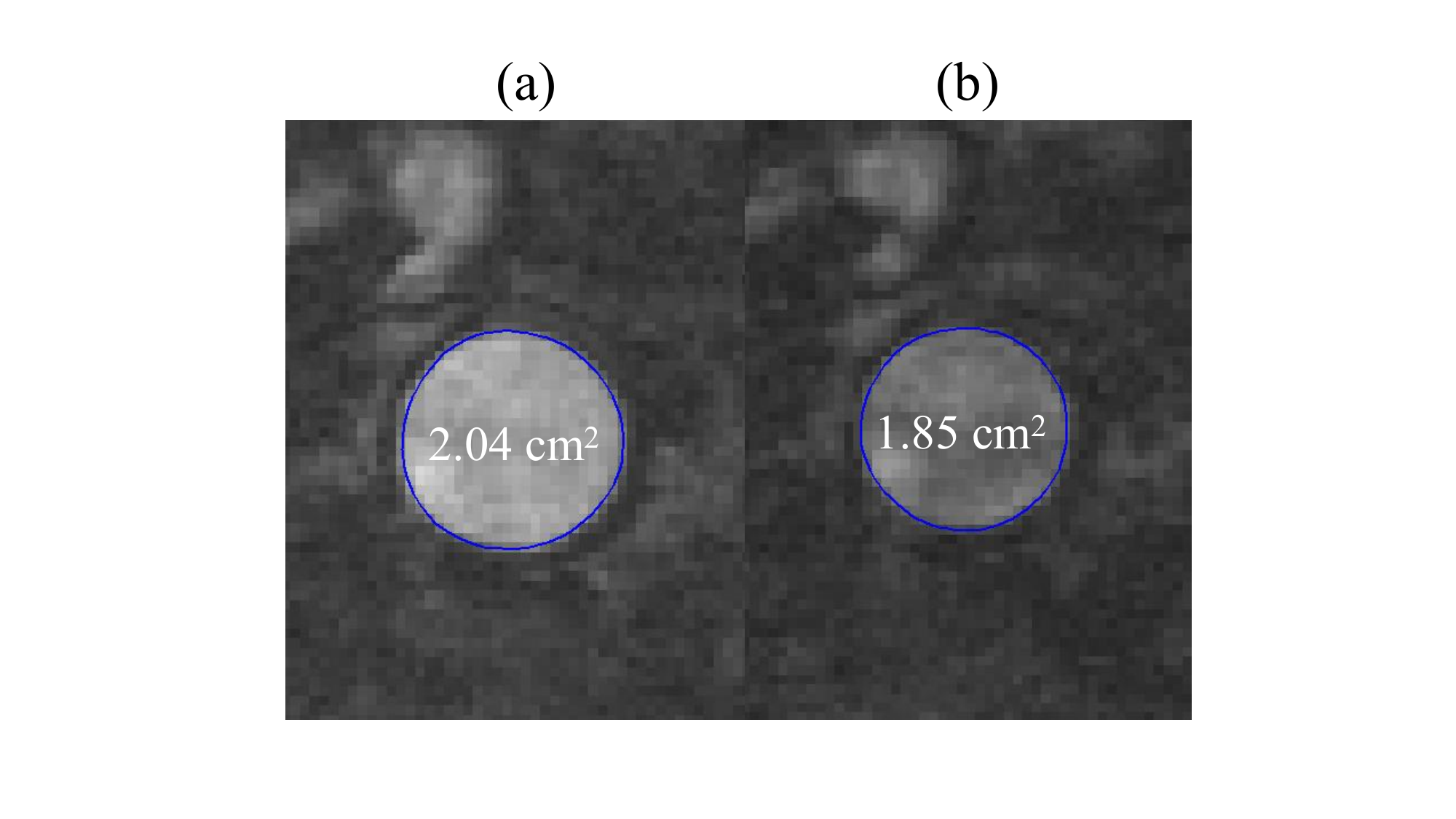}
		\end{center}
		\caption{Magnitude images of 2D PC-MRI at the moments of maximum (a) and minimum (b) cross-sectional areas.}
	\label{fig:PCMRI_result}
\end{figure}

\begin{table}[htbp]
	\begin{center}
		\tabcolsep=0.15cm
		\renewcommand{\arraystretch}{1.3}
		\begin{tabular}{ c c c c}
			\hline
			&Case & Area variation (cm$^2$)&Relative change in area \\
			\hline
			&1 & $1.85 \sim 1.96$ & 5.95\%\\
			&2 &$1.87 \sim 2.05$ & 9.63\% \\
			&3 & $1.92 \sim 2.04$& 6.25\% \\
			&4 &$1.92 \sim 2.02$ & 5.21\% \\
			&PC-MRI &$1.85 \sim 2.04$ & 10.27\% \\
			\hline
		\end{tabular}
	\end{center}
	\caption{A comparison of the area variations between PC-MRI measurements and FSI simulations at the slice indicated in Figure \ref{fig:AA_diagram1}.}
	\label{table:comparison_area_ariations}
\end{table}

Figure \ref{fig:PCMRI_result} presents the 2D PC-MRI images at the slice indicated in Figure \ref{fig:AA_diagram1}. The regions of the lumen are delineated by blue closed curves, with the numbers indicating the areas of the enclosed regions. The area is calculated by counting the number of pixels.  We further compare the area variations at this cross-section between the PC-MRI measurement and FSI simulations, with the results listed in Table \ref{table:comparison_area_ariations}. The area variation measured by PC-MRI is in close agreement with predictions from the GOH model of $\kappa = 0.226$. The rest models give an over-stiff behavior of the wall, making the relative area change significantly less than the measurement. This signifies that incorporating fiber dispersion within the GOH model offers a more accurate prediction of the behavior of the abdominal arterial wall. 

\section{Conclusion}
\label{sec:conclusion}
In this study, we first presented a systematic modeling pipeline for fibrous arterial walls in vascular FSI analysis. We proposed modeling the wall based on segmentation data and CAD tools. It is constructed upon the mature lumen modeling strategy, rendering it readily integrable into the mainstream cardiovascular modeling pipeline \cite{Arthurs2021,Updegrove2017}. In comparison with the tissue model generated via the extrusion procedure \cite{Marchandise2013,Raut2015,Wu2022}, the proposed strategy addresses the potential self-intersection issue for multi-branched geometries in a robust and flexible manner. Moreover, we have developed a specialized mesh generation procedure that allows generating BL mesh for the fluid subdomain and ensures mesh continuity across the luminal interface. Leveraging the centerline information, we introduced a morphology-based method to construct local basis vectors with the goal of facilitating the accurate description of fiber orientation in the image-based vascular geometric model. Specialized procedures have been developed to effectively handle thick-wall geometries, multi-branched vessels, and bifurcation regions. Compared with methods that utilized elastic stress field to define local basis vectors, our approach demonstrates improved quality in delivering local basis vectors.

We analyzed the biomechanical significance of anisotropy through two numerical examples. Our findings indicate that the wall deformation and stress of neo-Hookean and GOH wall models under physiological loading are drastically different. By adjusting the dispersion parameter and fiber orientation, we confirmed that the difference arises from anisotropy instead of the exponential energy form. In both examples, we also observed that anisotropic models tend to deliver a localized stress distribution, whereas the isotropic models develop smoother stress fields. We identified that the stress distribution in the GOH model correlates with $\mathcal I$, the term appearing in the exponential function of the free energy. Furthermore, a GOH model accounting for fiber dispersion shows high-quality agreement with PC-MRI measurements, underscoring the importance of physiologically detailed arterial wall models in patient-specific investigations. The results indicate that the impact of anisotropy is non-negligible when analyzing the biomechanical behavior of the wall. On the other hand, we also observed that hemodynamic factors, such as WSS and flow field, are less sensitive to the choice of the wall model. Therefore, it may be acceptable to adopt a simple isotropic model when the focus is solely on the blood flow.

Several areas for future research warrant further investigation. First, the proposed pipeline can be conveniently generalized to specify layered structure within the tissue, allowing the description of distinct material properties and fiber orientations in the media and adventitia layers. Future research will focus on developing models that accurately represent this layered structure and exploring the biomechanical behavior of such layered models, particularly in the FSI setting. Second, refining the models to better simulate fibers is essential. This includes considering out-of-plane dispersions and excluding fiber responses under compression. These refinements will contribute to a more accurate representation of the arterial wall and assist in better understanding its biomechanics and mechanobiology properties.

\section*{Acknowledgements}
This work is supported by the National Natural Science Foundation of China [Grant Numbers 12172160], Shenzhen Science and Technology Program [Grant Number JCYJ20220818100600002], Southern University of Science and Technology [Grant Number Y01326127], and the Department of Science and Technology of Guangdong Province [2021QN020642]. Computational resources are provided by the Center for Computational Science and Engineering at the Southern University of Science and Technology.

\section*{Ethical Statement}
Ethical statement is not applicable for this article.

\bibliographystyle{plain}
\bibliography{vascular-fsi}

\begin{thebibliography}{10}

\bibitem{segment-website}
{MEDVISO} {S}egment.
\newblock \url{https://medviso.com/segment}.

\bibitem{opencascade-reference}
Open {CASCADE} online documentation.
\newblock \url{https://dev.opencascade.org/doc/overview/html/index.html}.

\bibitem{vmtk-website}
The {V}ascular {M}odeling {T}ool{K}it.
\newblock \url{www.vmtk.org}.

\bibitem{Alastrue2010}
V.~Alastru{\'e}, A.~Gar{\'\i}a, E.~Pe{\~n}a, J.F. Rodr{\'\i}guez, M.A.
  Mart{\'\i}nez, and M.~Doblar{\'e}.
\newblock Numerical framework for patient-specific computational modelling of
  vascular tissue.
\newblock {\em International Journal for Numerical Methods in Biomedical
  Engineering}, 26(1):35--51, 2010.

\bibitem{Antiga2008}
L.~Antiga, M.~Piccinelli, L.~Botti, B.~Ene-Iordache, A.~Remuzzi, and D.A.
  Steinman.
\newblock An image-based modeling framework for patient-specific computational
  hemodynamics.
\newblock {\em Medical \& Biological Engineering \& Computing}, 46:1097, 2008.

\bibitem{Antiga2008a}
L.~Antiga, B.~A. Wasserman, and D.~A. Steinman.
\newblock On the overestimation of early wall thickening at the carotid bulb by
  black blood {MRI}, with implications for coronary and vulnerable plaque
  imaging.
\newblock {\em Magnetic Resonance in Medicine: An Official Journal of the
  International Society for Magnetic Resonance in Medicine}, 60(5):1020--1028,
  2008.

\bibitem{Arthurs2021}
C.J. Arthurs, R.~Khlebnikov, A.~Melville, M.~Mar{\v{c}}an, A.~Gomez,
  D.~Dillon-Murphy, F.~Cuomo, M.~Silva Vieira, J.~Schollenberger, S.R. Lynch,
  et~al.
\newblock {CRIMSON}: {A}n open-source software framework for cardiovascular
  integrated modelling and simulation.
\newblock {\em PLOS Computational Biology}, 17(5):e1008881, 2021.

\bibitem{Balzani2016}
D.~Balzani, S.~Deparis, S.~Fausten, D.~Forti, A.~Heinlein, A.~Klawonn,
  A.~Quarteroni, O.~Rheinbach, and J.~Schr{\"o}der.
\newblock Numerical modeling of fluid-structure interaction in arteries with
  anisotropic polyconvex hyperelastic and anisotropic viscoelastic material
  models at finite strains.
\newblock {\em International Journal for Numerical Methods in Biomedical
  Engineering}, 32(10):e02756, 2016.

\bibitem{Balzani2023}
D.~Balzani, A.~Heilein, A.~Klawonn, O.~Rheinbach, and J.~Schr{\"o}der.
\newblock Comparison of arterial wall models in fluid–structure interaction
  simulations.
\newblock {\em Computational Mechanics}, 72(5):949--965, 2023.

\bibitem{Balzani2006}
D.~Balzani, P.~Neff, J.~Schr{\"o}der, and G.A. Holzapfel.
\newblock A polyconvex framework for soft biological tissues. adjustment to
  experimental data.
\newblock {\em International Journal of Solids and Structures},
  43(20):6052--6070, 2006.

\bibitem{Baeumler2020}
K.~B\"{a}umler, V.~Vedula, A.M. Sailer, J.~Seo, P.~Chiu, G.~Mistelbauer, F.P.
  Chan, M.P. Fischbein, A.L. Marsden, and D.~Fleischmann.
\newblock Fluid-structure interaction simulations of patient-specific aortic
  dissection.
\newblock {\em Biomechanics and Modeling in Mechanobiology}, 19:1607--1628,
  2020.

\bibitem{Bazilevs2010}
Y.~Bazilevs, M.C. Hsu, Y.~Zhang, W.~Wang, T.~Kvamsdal, S.~Hentschel, and J.G.
  Isaksen.
\newblock Computational vascular fluid-structure interaction: methodology and
  application to cerebral aneurysms.
\newblock {\em Biomechanics and Modeling in Mechanobiology}, 9(4):481--498,
  2010.

\bibitem{Botsch2010}
M.~Botsch, L.~Kobbelt, M.~Pauly, P.~Alliez, and B.~L{\'e}vy.
\newblock {\em Polygon Mesh Processing}.
\newblock CRC press, 2010.

\bibitem{Brands2008}
D.~Brands, A.~Klawonn, O.~Rheinbach, and J.~Schr{\"o}der.
\newblock Modelling and convergence in arterial wall simulations using a
  parallel {FETI} solution strategy.
\newblock {\em Computer Methods in Biomechanics and Biomedical Engineering},
  11(5):569--583, 2008.

\bibitem{Bruneau2023}
D.A. Bruneau, K.~Valen-Sendstad, and D.A. Steinman.
\newblock Onset and nature of flow-induced vibrations in cerebral aneurysms via
  fluid–structure interaction simulations.
\newblock {\em Biomechanics and Modeling in Mechanobiology}, 22:761--771, 2023.

\bibitem{Bukac2019}
M.~Buka{\v{c}}, S.~{\v{C}}ani{\'c}, J.~Tamba{\v{c}}a, and Y.~Wang.
\newblock Fluid-structure interaction between pulsatile blood flow and a curved
  stented coronary artery on a beating heart: {A} four stent computational
  study.
\newblock {\em Computer Methods in Applied Mechanics and Engineering},
  350:679--700, 2019.

\bibitem{Caro2012}
C.G. Caro, T.J. Pedley, R.C. Schroter, and W.A. Seed.
\newblock {\em The mechanics of the circulation}.
\newblock Cambridge University Press, 2012.

\bibitem{Cenanovic2020}
M.~Cenanovic, P.~Hansbo, and M.G. Larson.
\newblock Finite element procedures for computing normals and mean curvature on
  triangulated surfaces and their use for mesh refinement.
\newblock {\em Computer Methods in Applied Mechanics and Engineering},
  372:113445, 2020.

\bibitem{Chagnon2015}
G.~Chagnon, M.~Rebouah, and D.~Favier.
\newblock Hyperelastic energy densities for soft biological tissues: a review.
\newblock {\em Journal of Elasticity}, 120:129--160, 2015.

\bibitem{Chan2001}
T.F. Chan and L.A. Vese.
\newblock Active contours without edges.
\newblock {\em IEEE Transactions on Image Processing}, 10(2):266--277, 2001.

\bibitem{Dalbosco2024}
M.~Dalbosco, E.A. Fancello, and G.A. Holzapfel.
\newblock Multiscale computational modeling of arterial micromechanics: {A}
  review.
\newblock {\em Computer Methods in Applied Mechanics and Engineering},
  425:116916, 2024.

\bibitem{Demiray1972}
H.~Demiray.
\newblock A note on the elasticity of soft biological tissues.
\newblock {\em Journal of Biomechanics}, 5:309--311, 1972.

\bibitem{Driessen2003}
N.J.B. Driessen, G.W.M. Peters, J.M. Huyghe, C.V.C. Bouten, and F.P.T.
  Baaijens.
\newblock Remodelling of continuously distributed collagen fibres in soft
  connective tissues.
\newblock {\em Journal of Biomechanics}, 36(8):1151--1158, 2003.

\bibitem{Figueroa2006}
C.A. Figueroa, I.E. Vignon-Clementel, K.E. Jansen, T.J.R. Hughes, and C.A.
  Taylor.
\newblock A coupled momentum method for modeling blood flow in
  three-dimensional deformable arteries.
\newblock {\em Computer Methods in Applied Mechanics and Engineering},
  195:5685--5706, 2006.

\bibitem{Garimella2000}
R.V. Garimella and M.S. Shephard.
\newblock Boundary layer mesh generation for viscous ow simulations.
\newblock {\em International Journal for Numerical Methods in Engineering},
  49:193--218, 2000.

\bibitem{Gasser2006}
T.C. Gasser, R.W. Ogden, and G.A. Holzapfel.
\newblock Hyperelastic modelling of arterial layers with distributed collagen
  fibre orientations.
\newblock {\em Journal of the Royal Society Interface}, 3(6):15--35, 2006.

\bibitem{Gee2010}
M.W. Gee, C.H. F\"{o}rster, and W.A. Wall.
\newblock A computational strategy for prestressing patient-specific
  biomechanical problems under finite deformation.
\newblock {\em International Journal for Numerical Methods in Biomedical
  Engineering}, 26:52--72, 2010.

\bibitem{Geuzaine2009}
C.~Geuzaine and J.F. Remacle.
\newblock Gmsh: {A} 3-{D} finite element mesh generator with built-in pre- and
  post-processing facilities.
\newblock {\em International Journal for Numerical Methods in Biomedical
  Engineering}, 79:1309--1331, 2009.

\bibitem{Hariton2007}
I.~Hariton, G.~Debotton, T.C. Gasser, and G.A. Holzapfel.
\newblock Stress-driven collagen fiber remodeling in arterial walls.
\newblock {\em Biomechanics and Modeling in Mechanobiology}, 6:163--175, 2007.

\bibitem{Hariton2007b}
I.~Hariton, G.~Debotton, T.C. Gasser, and G.A. Holzapfel.
\newblock Stress-modulated collagen fiber remodeling in a human carotid
  bifurcation.
\newblock {\em Journal of Theoretical Biology}, 248(3):460--470, 2007.

\bibitem{Hirschhorn2020}
M.~Hirschhorn, V.~Tchantchaleishvili, R.~Stevens, J.~Rossano, and
  A.~Throckmorton.
\newblock Fluid-structure interaction modeling in cardiovascular medicine - {A}
  systematic review 2017–2019.
\newblock {\em Medical Engineering \& Physics}, 78:1--13, 2020.

\bibitem{Holzapfel2000a}
G.A. Holzapfel.
\newblock {\em Nonlinear {S}olid {M}echanics: {A} {C}ontinuum {A}pproach for
  {E}ngineering}.
\newblock John Wiley \& Sons, 2000.

\bibitem{Holzapfel2006}
G.A. Holzapfel.
\newblock Determination of material models for arterial walls from uniaxial
  extension tests and histological structure.
\newblock {\em Journal of Theoretical Biology}, 238(2):290--302, 2006.

\bibitem{Holzapfel2000}
G.A. Holzapfel, T.C. Gasser, and R.W. Ogden.
\newblock A new constitutive framework for arterial wall mechanics and a
  comparative study of material models.
\newblock {\em Journal of Elasticity and the Physical Science of Solids},
  61:1--48, 2000.

\bibitem{Holzapfel2015a}
G.A. Holzapfel, J.A. Niestrawska, R.W. Ogden, A.J. Reinisch, and A.J. Schriefl.
\newblock Modelling non-symmetric collagen fibre dispersion in arterial walls.
\newblock {\em Journal of the Royal Society Interface}, 12(106):20150188, 2015.

\bibitem{Holzapfel2015}
G.A. Holzapfel and R.W. Ogden.
\newblock On the tension-compression switch in soft fibrous solids.
\newblock {\em European Journal of Mechanics-A/Solids}, 49:561--569, 2015.

\bibitem{Holzapfel2005}
G.A. Holzapfel, G.~Sommer, C.T. Gasser, and P.~Regitnig.
\newblock Determination of layer-specific mechanical properties of
  humancoronary arteries with nonatherosclerotic intimal thickeningand related
  constitutive modeling.
\newblock {\em American Journal of Physiology-Heart and Circulatory
  Physiology}, 289(5):H2048--H2058, 2005.

\bibitem{Holzapfel2004a}
G.A. Holzapfel, G.~Sommer, and P.~Regitnig.
\newblock Anisotropic {{Mechanical Properties}} of {{Tissue Components}} in
  {{Human Atherosclerotic Plaques}}.
\newblock {\em Journal of Biomechanical Engineering}, 126(5):657--665, 2004.

\bibitem{Hsu2011}
M.C. Hsu and Y.~Bazilevs.
\newblock Blood vessel tissue prestress modeling for vascular fluid–structure
  interaction simulation.
\newblock {\em Finite Elements in Analysis and Design}, 47(6):593--599, 2011.

\bibitem{Huang2023}
M.~Huang, D.~Tang, L.~Wang, R.~Lv, G.~Ma, and G.S. Mintz.
\newblock Comparison of multilayer and single-layer coronary plaque models on
  stress/strain calculations based on optical coherence tomography images.
\newblock {\em Frontiers in Physiology}, 14:1251401, 2023.

\bibitem{Humphrey2002}
J.D. Humphrey.
\newblock {\em Cardiovascular {S}olid {M}echanics: {C}ells, {T}issues, and
  {O}rgans}.
\newblock Springer, 2002.

\bibitem{Humphrey2021}
J.D. Humphrey and M.A. Schwartz.
\newblock Vascular mechanobiology: Homeostasis, adaptation,and disease.
\newblock {\em Annual Review of Biomedical Engineering}, 23:1--27, 2021.

\bibitem{Jin2005}
S.~Jin, R.R. Lewis, and D.~West.
\newblock A comparison of algorithms for vertex normal computation.
\newblock {\em The visual computer}, 21:71--82, 2005.

\bibitem{Kang2009}
W.X. Kang, Q.Q. Yang, and R.P. Liang.
\newblock The comparative research on image segmentation algorithms.
\newblock In {\em 2009 First International Workshop on Education Technology and
  Computer Science}, volume~2, pages 703--707. IEEE, 2009.

\bibitem{Kiousis2009}
D.E. Kiousis, S.F. Rubinigg, M.~Auer, and G.A. Holzapfel.
\newblock A methodology to analyze changes in lipid core and calcification onto
  fibrous cap vulnerability: the human atherosclerotic carotid bifurcation as
  an illustratory example.
\newblock {\em Journal of Biomechanical Engineering}, 131(12):121002, 2009.

\bibitem{Ladak2001}
H.M. Ladak, J.B. Thomas, J.R. Mitchell, B.K. Rutt, and D.A. Steinman.
\newblock A semi-automatic technique for measurement of arterial wall from
  black blood {MRI}.
\newblock {\em Medical Physics}, 28(6):1098--1107, 2001.

\bibitem{Lan2023}
I.S. Lan, J.~Liu, W.~Yang, J.~Zimmermann, D.B. Ennis, and A.L. Marsden.
\newblock Validation of the reduced unified continuum formulation against in
  vitro 4{D}-flow {MRI}.
\newblock {\em Annals of Biomedical Engineering}, 51:377--393, 2023.

\bibitem{Liu2018}
J.~Liu and A.L. Marsden.
\newblock A unified continuum and variational multiscale formulation for
  fluids, solids, and fluid-structure interaction.
\newblock {\em Computer Methods in Applied Mechanics and Engineering},
  337:549--597, 2018.

\bibitem{Liu2020}
J.~Liu, W.~Yang, I.S. Lan, and A.L. Marsden.
\newblock Fluid-structure interaction modeling of blood flow in the pulmonary
  arteries using the unified continuum and variational multiscale formulation.
\newblock {\em Mechanics Research Communications}, 107:103556, 2020.

\bibitem{Lopes2019}
D.~Lopes, H.~Puga, J.C. Teixeira, and S.F. Teixeira.
\newblock Influence of arterial mechanical properties on carotid blood flow:
  {C}omparison of {CFD} and {FSI} studies.
\newblock {\em International Journal of Mechanical Sciences}, 160:209--218,
  2019.

\bibitem{Marchandise2013}
E.~Marchandise, C.~Geuzaine, and J.F. Remacle.
\newblock Cardiovascular and lung mesh generation based on centerlines.
\newblock {\em International Journal for Numerical Methods in Biomedical
  Engineering}, 29:665--682, 2013.

\bibitem{Misiulis2019}
E.~Misiulis, A.~D{\v{z}}iugys, R.~Navakas, and V.~Petkus.
\newblock A comparative study of methods used to generate the arterial fiber
  structure in a clinically relevant numerical analysis.
\newblock {\em International Journal for Numerical Methods in Biomedical
  Engineering}, 35(6):e3194, 2019.

\bibitem{Mobadersany2023}
N.~Mobadersany, N.H. Meshram, P.~Kemper, S.V. Sise, G.M. Karageorgos, P.~Liang,
  G.A. Ateshian, and E.E. Konofagou.
\newblock Pulse wave imaging of a stenotic artery model with plaque
  constituents of different stiffnesses: {E}xperimental demonstration in
  phantoms and fluid-structure interaction simulation.
\newblock {\em Journal of Biomechanics}, 149:111502, 2023.

\bibitem{Moerman2022}
K.M. Moerman, P.~Konduri, B.~Fereidoonnezhad, H.~Marquering, A.~van~der Lugt,
  G.~Luraghi, S.~Bridio, F.~Migliavacca, J.F.R. Matas, and P.~McGarry.
\newblock Development of a patient-specific cerebral vasculature
  fluid-structure-interaction model.
\newblock {\em Journal of Biomechanics}, 133:110896, 2022.

\bibitem{Nama2020}
N.~Nama, M.~Aguirre, J.D. Humphrey, and C.A. Figueroa.
\newblock A nonlinear rotation-free shell formulation with prestressing for
  vascular biomechanics.
\newblock {\em Scientific Reports}, 10:17528, 2020.

\bibitem{Nolan2014}
D.R. Nolan, A.L. Gower, M.~Destrade, R.W. Ogden, and J.P. McGarry.
\newblock A robust anisotropic hyperelastic formulation for the modelling of
  soft tissue.
\newblock {\em Journal of the Mechanical Behavior of Biomedical Materials},
  39:48--60, 2014.

\bibitem{Chen2022}
C.~Peng, J.~Liu, W.~He, W.~Qin, T.~Yuan, Y.~Kan, K.~Wang, S.~Wang, and Y.~Shi.
\newblock Numerical simulation in the abdominal aorta and the visceral arteries
  with or without stenosis based on {{{\textsc{2D PCMRI}}}}.
\newblock {\em International Journal for Numerical Methods in Biomedical
  Engineering}, 38(3):e3569, 2022.

\bibitem{Piccinelli2009}
M.~Piccinelli, A.~Veneziani, D.A. Steinman, A.~Remuzzi, and L.~Antiga.
\newblock A {{Framework}} for {{Geometric Analysis}} of {{Vascular
  Structures}}: {{Application}} to {{Cerebral Aneurysms}}.
\newblock {\em IEEE Transactions on Medical Imaging}, 28(8):1141--1155, 2009.

\bibitem{Pukaluk2024}
A.~Pukaluk, G.~Sommer, and G.A. Holzapfel.
\newblock Multimodal experimental studies of the passive mechanical behavior of
  human aortas: current approaches and future directions.
\newblock {\em Acta Biomaterialia}, 178:1--12, 2024.

\bibitem{Raut2015}
S.S. Raut, P.~Liu, and E.A. Finol.
\newblock An approach for patient-specific multi-domain vascular mesh
  generation featuring spatially varying wall thickness modeling.
\newblock {\em Journal of Biomechanics}, 48:1972--1981, 2015.

\bibitem{Reymond2013}
P.~Reymond, P.~Crosetto, S.~Deparis, A.~Quarteroni, and N.~Stergiopulos.
\newblock Physiological simulation of blood flow in the aorta: {C}omparison of
  hemodynamic indices as predicted by 3-{D} {FSI}, 3-{D} rigid wall and 1-{D}
  models.
\newblock {\em Medical engineering \& physics}, 35(6):784--791, 2013.

\bibitem{Roy2014}
D.~Roy, G.A. Holzapfel, C.~Kauffmann, and G.~Soulez.
\newblock Finite element analysis of abdominal aortic aneurysms: geometrical
  and structural reconstruction with application of an anisotropic material
  model.
\newblock {\em The IMA Journal of Applied Mathematics}, 79(5):1011--1026, 2014.

\bibitem{Senthilkumaran2016}
N.~Senthilkumaran and S.~Vaithegi.
\newblock Image segmentation by using thresholding techniques for medical
  images.
\newblock {\em Computer Science \& Engineering: An International Journal},
  6(1):1--13, 2016.

\bibitem{Sherifova2019}
S.~Sherifova and G.A. Holzapfel.
\newblock Biomechanics of aortic wall failure with a focus on dissection and
  aneurysm: {A} review.
\newblock {\em Acta Biomaterialia}, 99:1--17, 2019.

\bibitem{Shi2021}
Y.~Shi, C.~Peng, J.~Liu, H.~Lan, C.~Li, W.~Qin, T.~Yuan, Y.~Kan, S.~Wang, and
  W.~Fu.
\newblock A modified method of computed fluid dynamics simulation in abdominal
  aorta and visceral arteries.
\newblock {\em Computer Methods in Biomechanics and Biomedical Engineering},
  24(15):1718--1729, 2021.

\bibitem{Si2015}
H.~Si.
\newblock {TetGen}, a {D}elaunay-based quality tetrahedral mesh generator.
\newblock {\em ACM Transactions on Mathematical Software}, 41:1--36, 2015.

\bibitem{Souche2022}
A.~Souche and K.~Valen-Sendstad.
\newblock High-fidelity fluid structure interaction simulations of
  turbulent-likeaneurysm flows reveals high-frequency narrowband wall
  vibrations: {A}stimulus of mechanobiological relevance?
\newblock {\em Journal of Biomechanics}, 145:111369, 2022.

\bibitem{Steinman2002}
D.A. Steinman, J.B. Thomas, H.M. Ladak, J.S. Milner, B.K. Rutt, and J.D.
  Spence.
\newblock Reconstruction of carotid bifurcation hemodynamics and wall thickness
  using computational fluid dynamics and {MRI}.
\newblock {\em Magnetic Resonance in Medicine}, 47:149--159, 2002.

\bibitem{Suito2014}
H.~Suito, K.~Takizawa, V.Q.H. Huynh, D.~Sze, and T.~Ueda.
\newblock {FSI} analysis of the blood flow and geometrical characteristics in
  the thoracic aorta.
\newblock {\em Computational Mechanics}, 54:1035--1045, 2014.

\bibitem{Sun2024}
Y.~Sun, Q.~Lu, and J.~Liu.
\newblock A parallel solver framework for fully implicit monolithic
  fluid-structure interaction.
\newblock {\em Acta Mechanica Sinica}, 40:324074, 2024.

\bibitem{Takizawa2010a}
K.~Takizawa, J.~Christopher, T.E. Tezduyar, and S.~Sathe.
\newblock Space–time finite element computation of arterial fluid–structure
  interactions with patient-specific data.
\newblock {\em International Journal for Numerical Methods in Biomedical
  Engineering}, 26:101--116, 2010.

\bibitem{Tang2004}
D.~Tang, C.~Yang, J.~Zheng, P.K. Woodard, G.A. Sicard, J.E. Saffitz, and
  C.~Yuan.
\newblock 3{D} {MRI}-based multicomponent {FSI} models for atherosclerotic
  plaques.
\newblock {\em Annals of biomedical engineering}, 32:947--960, 2004.

\bibitem{Taylor2013}
C.A. Taylor, T.A. Fonte, and J.K. Min.
\newblock Computational {F}luid {D}ynamics {A}pplied to {C}ardiac {C}omputed
  {T}omography for {N}oninvasive {Q}uantification of {F}ractional {F}low
  {R}eserve: {S}cientific {B}asis.
\newblock {\em Journal of the American College of Cardiology}, 61:2233--2241,
  2013.

\bibitem{Taylor2023}
C.A. Taylor, K.~Petersen, N.~Xiao, M.~Sinclair, Y.~Bai, S.R. Lynch, A.~UpdePac,
  and M.~Schaap.
\newblock Patient-specific modeling of blood flow in the coronary arteries.
\newblock {\em Computer Methods in Applied Mechanics and Engineering},
  417:116414, 2023.

\bibitem{Tricerri2015}
P.~Tricerri, L.~Ded{\`e}, S.~Deparis, A.~Quarteroni, A.M. Robertson, and
  A.~Sequeira.
\newblock Fluid-structure interaction simulations of cerebral arteries modeled
  by isotropic and anisotropic constitutive laws.
\newblock {\em Computational Mechanics}, 55:479--498, 2015.

\bibitem{Updegrove2017}
A.~Updegrove, N.M. Wilson, J.~Merkow, H.~Lan, A.L. Marsden, and S.C. Shadden.
\newblock Sim{V}ascular: {A}n {O}pen {S}ource {P}ipeline for {C}ardiovascular
  {S}imulation.
\newblock {\em Annals of Biomedical Engineering}, 45:525--541, 2017.

\bibitem{Updegrove2016}
A.~Updegrove, N.M. Wilson, and S.C. Shadden.
\newblock Boolean and smoothing of discrete polygonal surfaces.
\newblock {\em Advances in Engineering Software}, 95:16--27, 2016.

\bibitem{Vese2002}
L.A. Vese and T.F. Chan.
\newblock A multiphase level set framework for image segmentation using the
  mumford and shah model.
\newblock {\em International Journal of Computer Vision}, 50:271--293, 2002.

\bibitem{Wu2022}
C.~Wu, X.~Liu, D.~Ghista, Y.~Yin, and H.~Zhang.
\newblock Effect of plaque compositions on fractional fow reserve in a
  fuid-structure interaction analysis.
\newblock {\em Biomechanics and Modeling in Mechanobiology}, 21:203--220, 2022.

\bibitem{Xiang2020}
Y.~Xiang, D.~Zhong, S.~Rudykh, H.~Zhou, S.~Qu, and W.~Yang.
\newblock A review of physically based and thermodynamically based constitutive
  models for soft materials.
\newblock {\em Journal of Applied Mechanics}, 87(11):110801, 2020.

\bibitem{Yang2009}
C.~Yang, R.G. Bach, J.~Zheng, I.E. Naqa, P.K. Woodard, Z.~Teng, K.~Billiar, and
  D.~Tang.
\newblock In vivo {IVUS}-based 3-{D} fluid--structure interaction models with
  cyclic bending and anisotropic vessel properties for human atherosclerotic
  coronary plaque mechanical analysis.
\newblock {\em IEEE Transactions on biomedical engineering}, 56(10):2420--2428,
  2009.

\end{thebibliography}

\end{document}